\input harvmac.tex

\input epsf
\input tables.tex
\noblackbox
\nref\one{C. Bachas, ``Relativistic String in a Pulse,'' hep-th/0212217.}
\nref\two{B. Durin and B. Pioline, ``Open strings in relativistic ion traps,''
hep-th/0302159.}
\nref\three{Y. Hikida, H. Takayanagi, and T. Takayanagi, ``Boundary States 
for D-branes with Traveling Waves,'' hep-th/0303214.}
\nref\four{T. Vachaspati, ``Gravitational Effects of Cosmic Strings,''
Nucl. Phys. {\bf B277} (1986) 593\semi D. Garfinkle and T. Vachaspati,
``Cosmic String Travelling Waves,'' Phys. Rev. {\bf D42} (1990) 1960.}
\nref\five{G. Horowitz and A. Tseytlin, ``A New Class of Exact Solutions 
in String Theory,'' Phys. Rev. {\bf D51} (1995) 2896, 
hep-th/9409021\semi C. Callan, J. Maldacena, and 
A. Peet, ``Extremal Black Holes 
As Fundamental Strings,'' Nucl. Phys.
{\bf B475} (1996) 645, hep-th/9510134\semi A. Dabholkar, 
J. Gauntlett, J. Harvey, and D. Waldram,
``Strings as Solitons \& Black Holes as Strings,'' Nucl. Phys.
{\bf B474} (1996) 85, hep-th/9511053\semi O. Lunin and S. Mathur, ``Metric 
of the multiply wound rotating string,'' Nucl. Phys.
{\bf B610} (2001) 49, hep-th/0105136.} 
\nref\six{D. Mateos, S. Ng, and P. Townsend, ``Supercurves,'' Phys. Lett. 
{\bf B538} (2002) 366,
hep-th/0204062.}
\nref\seven{R. Myers and D. Winters, ``From D-Dbar Pairs to Branes in Motion,''
JHEP 0212 (2002) 061, hep-th/0211042.}
\nref\eight{J. Blum, ``Triple Intersections and Geometric Transitions,''
Nucl. Phys. {\bf B646} (2002) 524, hep-th/0207016.}
\nref\nine{F. Hussain, R. Iengo, C. N{\' u}{\~ n}ez, and C. Scrucca,``Closed 
string radiation from moving D-branes,'' Nucl. Phys.
{\bf B517} (1998) 92, hep-th/9710049.}
\nref\ten{C. Lovelace, Phys. Lett. 
{\bf 34B} (1971) 500\semi E. Cremmer and J. Scherk, Nucl. Phys.
{\bf B50} (1972) 222\semi M. Ademollo, A. D'Adda, R. D'Auria, F. Gliozzi,
E. Napolitano, S. Sciuto, and P. Di Vecchia, Nucl. Phys.
{\bf B77} (1974) 89; Nucl. Phys. {\bf B94} (1975).}
\nref\eleven{C. Callan, C. Lovelace, C. Nappi, and S. Yost, Nucl. Phys.
{\bf B293} (1987) 83; Nucl. Phys.
{\bf B308} (1988) 221; J. Polchinski and Y. Cai, Nucl. Phys.
{\bf B296} (1988) 91; P. Di Vecchia, M. Frau, I. Pesando, 
S. Sciuto, A. Lerda, and R. Russo, Nucl. Phys.
{\bf B507} (1997) 259, hep-th/9707068; P. Di Vecchia, M. Frau, A. Lerda,
and A. Liccardo, Nucl. Phys. {\bf B565} 
(2000) 397, hep-th/9906214.}
\nref\twelve{M. Sakellariadou, ``Gravitational waves emitted from 
infinite strings,'' Phys. Rev. {\bf D42} (1990) 354.}
\nref\thirteen{I. Gradshteyn and I. Ryzhik, ``Table of Integrals, Series,
and Products,'' Academic Press (1965).}

\def\tx{\vbox{\sl\centerline{Physics Department}%
\centerline{University of Texas at Austin}%
\centerline{Austin, TX 78712 USA}}}

\Title{\vbox{\baselineskip12pt
\hbox{UTTG-01-03}\hbox{hep-th/0304173}}}
{\vbox{\centerline{Gravitational Radiation from Travelling Waves on D-Strings}}}

{\bigskip
\centerline{Julie D. Blum}
\bigskip
\tx

\bigskip
\medskip
\centerline{\bf Abstract}
Boundary states that preserve supersymmetry are constructed for 
fractional D-strings with travelling waves on a ${\bf C}^3/
{{\bf Z}_2\times {\bf Z}_2}$ orbifold.  The gravitational
radiation emitted between two D-strings with antiparallel travelling waves
is calculated.  
}

\Date{7/03}

\newsec{Introduction}

Topological defects such as cosmic strings are not thought to be the
primary source of large scale structure in the universe while evidence
for some sort of inflationary scenario is compelling.  Nevertheless, they may 
play a lesser role in structure formation and other high energy
phenomena like baryogenesis, very energetic cosmic rays, and gamma ray
bursts.  These defects are predicted by string theory and possible
observable effects are worth considering.

Recent work has studied unidirectional travelling waves on 
Dirichlet(D)-strings as an exactly solvable two-dimensional conformal
field theory \one .  Purely left or right moving waves of arbitrary
profile on D-strings preserve one-quarter of the supersymmetry
in supersymmetric string theory.  These D-strings are dual under the
action of type IIB $SL(2,{\bf Z})$ S-duality to fundamental string
states with only left movers or right movers excited.  Here, we will
calculate the gravitational radiation emitted from the interaction of
two fractional D-strings with pulses moving in opposite directions on an $N=2$ 
supersymmetric ${\bf C}^3/{\bf Z}_2\times {\bf Z}_2$ orbifold of type IIB
string theory.  The boundary states in this calculation are 
supersymmetric although the interaction breaks supersymmetry and
allows for nontrivial radiation.  The interaction of two D-strings with 
pulses moving in the same direction preserves supersymmetry and does not 
radiate.  In this model the D-string does
not produce a conical deficit in the geometry due to the noncompact extra
dimensions.  Treating more ``realistic'' four-dimensional models seems
difficult, at least in type IIB, because the supergravity with
travelling wave requires that the Minkowski dimensions be 
asymptotically flat far from the D-strings.

In the next section we show that the supergravity travelling wave solution
can be applied to the case of a regular (unwrapped) D-string on the conifold.
A supergravity solution of purely fractional branes on the conifold that 
is asymptotically flat in the Minkowski dimensions has yet to be 
constructed but should exist.  Our calculation considers fractional
branes in order to restrict movement in the extra dimensions.  In
section three we construct the relevant boundary states and calculate
amplitudes with no insertions, in section four we calculate the 
gravitational radiation (dilatonic and axionic radiation are
included) from the nonsupersymmetric interaction, and
in section five we conclude.  
While this work was progressing there was
a paper \two\ discussing various T-dual phenomena related to the 
travelling wave.  As my work was nearing completion a second related
paper \three\ appeared which has some overlap with section three.  This paper
constructs boundary states for D-strings with travelling waves in the
bosonic string and calculates the amplitude between strings with oppositely
directed pulses.

\newsec{Supergravity Solution of a Wiggly D-string on the Conifold}

Exact solutions in nonsupersymmetric gravitational theories of travelling
waves on cosmic strings were constructed in \four .  These solutions and
generalizations could also be constructed in string theory or supergravity
\five .  Recent work has analyzed interesting configurations related
by various dualities to the travelling wave \six\seven .

In this paper we will calculate the amplitude of interactions of 
fractional branes on the ${\bf C}^3/{\bf Z}_2\times {\bf Z}_2$ orbifold.
The resolution of this orbifold contains three ${\bf P}^1$'s that
each look locally like the conifold.  By tuning two of the Kahler
parameters to be large, we can approximate the geometry of the conifold.
In the resolved orbifold the fractional branes are equivalent to bound
states of D3-branes wrapped on a conifold and a D1-brane.  Of course,
the string theory calculation can only be done in the orbifold limit.
In this limit there are three complex plane singularities.  To
avoid singularities of the metric away from the branes in the
orbifold dimensions, we assume that a small resolution of all
three ${\bf P}^1$'s does not dramatically affect the string
theory result.
Far from the branes the geometry should be little affected by whether
the branes are fractional or regular.   Thus, asymptotically 
the geometry is the conifold (or any
noncompact Calabi-Yau threefold) and Minkowski space.  

We consider a Dirichlet onebrane parallel to the lightcone directions 
$x^{\pm}=
{t\pm x^1\over \sqrt{2}}$.  The two transverse directions of Minkowski 
space we will denote by $\vec{y}$.  The string is localized at $\vec{y}=0$
and at the origin of the conifold ($r_{con}=0$) where $r_{con}$ is a radial
coordinate on the resolved conifold.  Adding the travelling wave amounts
to the $x^-$ dependent translation $\vec{y}=\vec{Y}(x^-)$ and requiring
the metric to be asymptotically flat in the Minkowski directions.  
The solution takes the form
\eqn\metric{\eqalign{ds^2 &=e^{-{\phi\over 2}}g_{\mu\nu}dx^{\mu}dx^{\nu}\cr 
&=-2 
e^{-2\phi}dx^+ dx^- 
-(e^{-2\phi}-1)
\|\dot{\vec{Y}}\|^2 {dx^-}^2 +2(e^{-2\phi}-1) 
\dot{\vec{Y}}\cdot d{\vec y} dx^- +{\| d\vec{y}\|}^2 +ds^2_
{con}\cr B_{+-}^{RR}&=1-e^{-2\phi}\cr B_{-i}^{RR}&=
\sqrt{2}\dot{Y}_i(e^{-2\phi}-1).\cr }}
Here, $\vec{Y}=(Y^2 , Y^3)$ depends only on $x^-$, the dot denotes
differentiation with respect to $x^-$, 
${1\over\sqrt{g}}\partial_{\mu}(\sqrt{g} g^{\mu\nu}\partial_{\nu} 
e^{2\phi})=-Q\delta^{(8)}({\vec x}-{\vec Y})$
where $\phi(\vec{y}',r_{con})$ is
the dilaton, $Q=64\pi^6{\alpha '}^3 g_s$, and $g_s$ is constant.  The
coordinate $\vec{y}'=\vec{y}-\vec{Y}(x^-)$.  For large distances in
the variable $R=\sqrt{\|\vec{y}'\|^2 +r^2_{con}}$ and no angular momentum,
$e^{2\phi}\sim 1+{Q\over 6\Omega_7 R^6}$ where 
$\Omega_7={\pi^4\over 3}$.  In the above formulas what we have called
$e^{\phi}$ for simplicity of notation includes a factor $1\over g_s$.
Supersymmetry is broken to $1\over 16$ by the conditions 
\eqn\susy{\eqalign{\gamma^{+-}\epsilon&=\epsilon^*\cr \gamma^-\epsilon&=0\cr
D_{\mu}^{(conifold)}\epsilon&=0.\cr }}
Here, $\epsilon$ is a supersymmetric spinor of type IIB supergravity.  
The effective theory has two
supercharges of negative chirality.  There is a T-dual configuration
with three fivebranes intersecting on a string (two fivebranes
intersect in a threebrane while the third fivebrane intersects one
of the two in a threebrane and the other in a string) with momentum in
eleven dimensional supergravity.  This triple intersection is different
from the one analyzed in \eight\ as it is nonchiral. \foot{Note that the
conjectured theories in \eight\ with $(0,2)$ supersymmetry obtained by wrapping
fivebranes on fourcycles in Calabi-Yau fourfolds have anomalies that
cannot be cancelled and are therefore inconsistent.}

\newsec{Boundary States and Interactions for Travelling Waves}

In this section we construct boundary states for the fractional D-string 
with travelling wave on the orbifold ${\bf C}^3/{\bf Z}_2\times {\bf Z}_2$.
Most of the ingredients needed for our calculation in this section and
the next can be found in \nine .  The boundary state formalism was
introduced in \ten\ as an expression of the duality between one loop 
open string and tree closed string amplitudes and developed in
\eleven\ as a solution of a boundary condition on left and right movers.

The four fractional branes satisfying open-closed duality constraints on
this orbifold take the form
\eqn\frbrane{|B_{\alpha}>=|F_0>+\sum_j\epsilon_{\alpha j}|F_j>}
where $\prod_j{\epsilon_{\alpha j}}=1$, 
$\epsilon_{\alpha j}=\pm 1$, $\alpha\in\{0,1,2,3\}$,
$j\in\{1,2,3\}$, $F_0$ corresponds to a onebrane in flat ten
dimensions, and $F_i$ corresponds to a fractional onebrane at one of the three
possible ${\bf C}^2/{\bf Z}_2$ orbifolds.  The three $|B_{\alpha}>$ with
$\alpha\neq 0$ are equivalent in the resolution of the orbifold to a
bound state of two threebranes wrapped on one of the three conifold
${\bf P}^1$'s, a threebrane and antithreebrane on the other two
conifold ${\bf P}^1$'s, plus a onebrane.  The state $|B_0>$ has two
threebranes wrapped on each conifold plus a onebrane.

The interaction of two boundary states is given by a cylinder amplitude 
in which a closed string is exchanged between the two boundaries.  The
length of the cylinder is $l$, and the amplitude takes the form
\eqn\amp{{\cal A}^{ab}_{\alpha\beta}=
\int_0^\infty dl <B_{\alpha},
\vec{Y}_{1a}|e^{-lH}|B_{\beta},\vec{Y}_{2b}>}
where $H$ is the closed string Hamiltonian and $\vec{Y}_{ia}$ 
are the amplitudes
for the travelling waves on the D-strings.  $\vec{Y}_{ia}=\vec{Y}_i(x^a)$
where $a=\pm$.  To construct the boundary
states, let us write the expansions for closed string bosonic and
fermionic fields as a function of $z=\sigma +i\tau$ where $\sigma$ is
periodic with $0\leq\sigma\leq 1$, $0\leq\tau\leq l$, and 
$\alpha'={1\over 2\pi}$.
\eqn\expansion{\eqalign{X^{\mu}(z)&={{\hat x^{\mu}}\over 2} -{z\over 2}
{\hat p^{\mu}}+{i\over\sqrt{4\pi}}\sum_{n\neq 0}{1\over n}a^{\mu}_{n}
e^{2\pi inz}\cr {\tilde X}^{\mu}({\bar z})&={{\hat x^{\mu}}\over 2} +{{\bar z}
\over 2}
{\hat p^{\mu}}+{i\over\sqrt{4\pi}}\sum_{n\neq 0}{1\over n}{\tilde a}^{\mu}_{n}
e^{-2\pi in{\bar z}}\cr Z^i (z)&={i\over\sqrt{4\pi}}\sum_r{1\over r}
b^i_{r}e^{2\pi irz}\cr {\tilde Z}^i ({\bar z})&={i\over\sqrt{4\pi}}
\sum_r{1\over r}
{\tilde b}^i_{r}e^{-2\pi ir{\bar z}}\cr
\Psi^{\mu}(z)&=\sum_p \psi^{\mu}_p e^{2\pi ipz}\cr
{\tilde \Psi}^{\mu}({\bar z})&=\sum_p {\tilde\psi}^{\mu}_p 
e^{-2\pi ip{\bar z}}\cr \Lambda^i (z)&=\sum_p\lambda^i_{p+{1\over 2}} 
e^{2\pi i(p+{1\over 2})z}\cr
{\tilde\Lambda}^i ({\bar z})&=\sum_p{\tilde\lambda}^i_{p+{1\over 2}} 
e^{-2\pi i(p+{1\over 2}){\bar z}}\cr}}
Here, $X^{\mu}(\Psi^{\mu})$ are bosonic (fermionic) coordinates in the
untwisted directions while $Z^i(\Lambda^i)$ are coordinates in the directions
twisted by a ${\bf Z}_2$ action, and $n\in {\bf Z}$, $r\in {\bf Z}+{1\over 2}$,
while $p\in {\bf Z}$ or ${\bf Z}+{1\over 2}$ depending on whether the 
fermion sector is Ramond (R) or Neveu-Schwarz (NS).  The commutation
relations are 
\eqn\commutation{\eqalign{[a^{\mu}_{n},a^{\nu}_{m}]&=n 
\eta^{\mu\nu}\delta_{n+m} \cr \{\psi^{\mu}_p,\psi^{\nu}_q\}&=\eta^{\mu\nu}
\delta_{p+q}\cr [b^i_r,{\bar b}_s^j]&=r\delta^{ij}\delta_{r+s},\cr}}
etc. where $\eta^{00}=-1$ and $\eta^{ij}=\delta^{ij}$ for
$i,j\neq 0$.

The boundary interaction action of the travelling wave at $\tau=0$
takes the form
\eqn\action{S_{+,\eta}=
\oint d\sigma\{{\vec Y}(X^+)\cdot\partial_{\tau}{\vec X}+
{1\over 2}\partial_+{\vec Y}(X^+)\cdot{\vec\Psi}_+\Psi^+_-\}}
where $X^{\mu}=X^{\mu}(z)+{\tilde X}^{\mu}({\bar z})$,
$\Psi^{\mu}_{\pm}\equiv\Psi^{\mu}(z)\pm i\eta{\tilde\Psi}^{\mu}({\bar z})$,
$\Lambda^i_{\pm}\equiv\Lambda^i (z)\pm i\eta{\tilde\Lambda}({\bar z})$,
and $\eta=\pm 1$.

The action is invariant under the supersymmetries $\epsilon$ with
\eqn\strans{\eqalign{\delta_{\epsilon}X^{+}&=\epsilon\Psi_-^{+}\cr
\delta_{\epsilon}{\vec X}&=\epsilon{\vec\Psi}_-\cr
\delta_{\epsilon}\Psi_-^+&=-2i\epsilon\partial_{\sigma}X^+\cr
\delta_{\epsilon}{\vec\Psi}_+&=-2\epsilon\partial_{\tau}{\vec X}.\cr}}
The boundary equations of motion yield
\eqn\motionbos{\eqalign{\partial_{\tau}X^{+}|_{boundary}&=0\cr
\partial_{\tau}X^{-}-\partial_+{\vec Y}\cdot\partial_{\tau}{\vec X}
-{1\over 2}\partial_+^2{\vec Y}\cdot{\vec\Psi}_+\Psi^+_-
|_{boundary}&=0\cr
\partial_{\sigma}{\vec X}-\partial_+{\vec Y}\partial_{\sigma}X^+
|_{boundary}&=0\cr
\partial_{\sigma}Z^i|_{boundary}&=0\cr
\Psi^+_+|_{boundary}&=0\cr
\Psi^-_+-\partial_+{\vec Y}{\vec\Psi}_+|_{boundary}&=0\cr
{\vec\Psi}_--\partial_+{\vec Y}\Psi^+_-|_{boundary}&=0\cr
\Lambda^i_-|_{boundary}&=0.\cr}}

To expand these equations in oscillator modes, let us assume that we
can expand ${\vec Y}$ and $\partial_+{\vec Y}$ in the form
\eqn\yexpand{\eqalign{Y^i(X^+(\sigma))&=\sum_{n\in {\bf Z}}
f^i_{-n} e^{2\pi in\sigma}\cr
\partial_+Y^i(X^+(\sigma))&=\sum_{n\in {\bf Z}}g^i_{-n} e^{2\pi in\sigma}\cr}}
where $f^i_{-n}$ and $g^i_{-n}$ are functions of the commuting modes $a^+_n$,
${\tilde a}^+_n$, and ${\hat x}^+$ of the form
\eqn\expansosc{f^i_{n}=\sum_{\sum_i p_i -\sum_j q_j=n}
c^i_{{\vec p},{\vec q}}({\hat x}^+)a^+_{p_1}\cdots a^+_{p_l}
{\tilde a}^+_{q_1}\cdots{\tilde a}^+_{q_m},}
$p_i , q_j$ are negative integers, and $n\neq 0$.  These
operators are well defined quantum operators
without ordering ambiguities, and $p^+=0$ 
at the boundary.  Note that 
${\vec f}_0={\vec Y}({\hat x}^+)+\cdots$ and 
${\vec g}_0=\partial_+{\vec Y}({\hat x}^+)+\cdots$.  We also use
the relations of oscillators below to make the $p_i$ and $q_j$
negative integers.

The boundary conditions for oscillator modes at $\tau=0$ are the following:
\eqn\bc{\eqalign{(a_n^{+}+{\tilde a}_{-n}^{+})|F_1 ,{\vec Y}_+,\eta>&=0\cr
(a_n^{-}+{\tilde a}_{-n}^{-}-\sum_m{\vec g}_{-m}\cdot({\vec a}_{n+m}
+{\tilde{\vec a}}_{-n-m})-{1\over\sqrt{\pi}}{\vec g}_n\cdot{\hat{\vec p}}
+O(\partial_+^2{\vec Y}))
|F_1 ,{\vec Y}_+,\eta>&=0\cr
(a_n^{\mu}-{\tilde a}_{-n}^{\mu}-\sum_m{\vec g}^i_{-m}\delta_i^{\mu}(a_{n+m}^+ 
-{\tilde a}_{-n-m}^+))|F_1 ,{\vec Y}_+,\eta>&=0\cr
(b_r^{i}-{\tilde b}_{-r}^{i})|F_1 ,{\vec Y}_+,\eta>&=0\cr
(\psi_p^{+}+i\eta{\tilde \psi}_{-p}^{+})
|F_1 ,{\vec Y}_+,\eta>&=0\cr
(\psi_p^{-}+i\eta{\tilde \psi}_{-p}^{-}
-\sum_m {\vec g}_{-m}\cdot({\vec\psi}_{p+m} +i\eta{\tilde{\vec\psi}}_{-p-m}))
|F_1 ,{\vec Y}_+,\eta>&=0\cr
(\psi_p^{\mu}-i\eta{\tilde \psi}_{-p}^{\mu}-\sum_m {\vec g}_{-m}(\psi^+_{p+m}
-i\eta{\tilde\psi}^+_{-p-m})
|F_1 ,{\vec Y}_+,\eta>&=0\cr
(\lambda_{p+{1\over 2}}^{i}-i\eta{\tilde 
\lambda}_{-p-{1\over 2}}^{i})|F_1 ,{\vec Y}_+,\eta>&=0\cr}}
for $\mu\in\{2,3,4,5\}$, $i\in\{2,3\}$, and $n\neq 0$.  Terms with $g_n$ and 
$n\neq 0$ are also of order $\partial_+^2{\vec Y}$.
For bosonic zero modes we have the conditions:
\eqn\bczm{\eqalign{{\hat p}^+|F_1 ,{\vec Y}_+,\eta>&=0\cr
({\hat p}^{-}-{\vec g}_0\cdot{\hat{\vec p}}-\sqrt{\pi}\sum_{m\neq 0}
{\vec g}_{-m}\cdot({\vec a}_m-{\tilde{\vec a}}_{-m})+O(\partial_+^2{\vec Y}))
|F_1 ,{\vec Y}_+,\eta>&=0\cr
({\hat x}^{\mu}-f^i_0\delta_i^{\mu})|F_1 ,{\vec Y}_+,\eta>&=0\cr}}
We have verified to order $\partial_+{\vec Y}$
that the boundary state satisfying the above equations
is invariant under a linear combination of the left and right
moving local supersymmetries as well as the Virasoro constraints which
imply conformal invariance.
Note that translational invariance in one of the light cone directions
is broken, and global supersymmetry, accordingly, is broken in half.  
The boundary state satisfying the above constraints can be
written as
\eqn\bostate{|F_1,{\vec Y}_+,\eta>
=\int dx^+ dx^-
\int {d^4 q\over(2\pi)^4}{\cal O}({\vec Y}_{+,\eta})|F_1,\eta>_{osc}\otimes
|F_1,\eta>_0\otimes |ghost>}
\eqn\boosc{\eqalign{|F_1,\eta>_{osc}&=exp[\sum_{n>0}{1\over n}(a_{-n}^-
{\tilde a}^+_{-n}+a_{-n}^+
{\tilde a}^-_{-n}+a_{-n}^{\mu}\cdot
{\tilde a}^{\mu}_{-n})\cr 
&+\sum_{r>0}{1\over r}(b_{-r}^{i}\cdot
{\tilde{\bar b}}^{i}_{-r} +{\tilde b}^{i}_{-r}\cdot{\bar b}^{i}_{-r})\cr &+
i\eta\sum_{p>0}({\psi}^-_{-p}{\tilde\psi}^+_{-p}+\psi^+_{-p}{\tilde\psi}^-_{-p}
+\psi_{-p}^{\mu}\cdot{\tilde\psi}^{\mu}_{-p}\cr 
&+\lambda_{-p+{1\over 2}}^{i}\cdot
{\tilde{\bar \lambda}}^{i}_{-p+{1\over 2}} 
+{\tilde \lambda}^{i}_{-p+{1\over 2}}\cdot
{\bar \lambda}^{i}_{-p+{1\over 2}})]|0>\cr}}
\eqn\bozer{|F_1,\eta>_0^{RR}=
exp[i\eta(\psi^-_0{\tilde\psi}_0^+ 
+{\bar\psi}^2{\tilde\psi}^2+{\bar\psi}^4{\tilde\psi}^4)]
(|0>\otimes|{\tilde 0}>_{{\vec q},x^{\pm}}).}
\eqn\opwave{{\cal O}({\vec Y}_{+,\eta})=
Pe^{\oint d\sigma\{{\vec Y}(X^+)\cdot\partial_{\tau}{\vec X}+
{1\over 2}\partial_+{\vec Y}(X^+)\cdot{\vec\Psi}_+\Psi^+_-\}}.}
The oscillator vacuum is annihilated by modes with $n,p,r>0$.  In the 
NS-NS sector there are additional fermion zero modes for the $\lambda^i$, but
the $\psi^{\mu}$ zero modes are not present.  Here, $\psi^{\mu}=
{-i \psi^{\mu}_0 +\psi^{\mu +1}_0\over\sqrt{2}}$ for $\mu\in\{2,4\}$.

The zero mode vacuum in the RR sector is defined by the conditions:
\eqn\zmv{\eqalign{{\hat x}^{\pm}(|0>\otimes|{\tilde 0}>_{{\vec q},x^{\pm}})
&=x^{\pm}\cr
{\hat {\vec p}}(|0>\otimes|{\tilde 0}>_{{\vec q},x^{\pm}})&={\vec q}\cr
\psi^+_0|0>={\tilde\psi}^{-}_0|{\tilde 0}>&=0\cr
\psi^{\mu}| 0>={\tilde{\bar \psi}}^{\mu}|{\tilde 0}>&=0.\cr}}
Interchanging ${\cal O}({\vec Y}_{+,\eta})$ and ${\cal O}({\vec Y}_{-,\eta})$ 
gives the boundary state for the ${\vec Y}(x^-)$
pulse.  We will not write the ghost contribution explicitly.  The ghosts
will cancel the contribution to the partition function from two 
untwisted directions.  

The full boundary state $|F_{\alpha}, {\vec Y}_+>$ satisfying a GSO projection
can be written as
\eqn\fullbo{\eqalign{|F_{i}, {\vec{Y}}_+>&=c_1(|F_{i}, {\vec{Y}}_+ ,+>_{NS-NS}+
|F_{i}, {\vec{Y}}_+ ,->_{NS-NS}+|F_{i}, {\vec{Y}}_+ ,+>_{RR}+
|F_{i}, {\vec{Y}}_+ ,->_{RR})\cr
|F_{0}, {\vec{Y}}_+>&=c_2(|F_{0}, {\vec{Y}}_+ ,+>_{NS-NS}-
|F_{0}, {\vec{Y}}_+ ,->_{NS-NS}+|F_{0}, {\vec{Y}}_+ ,+>_{RR}+
|F_{0}, {\vec{Y}}_+ ,->_{RR})\cr}}
up to overall normalizations $c_1$, $c_2$.  The closed string Hamiltonian is 
\eqn\ham{\eqalign{H_{\alpha}^s&=
{{\hat p}^2\over 2}+2\pi[\sum_{n>0}(-a^+_{-n} a^-_n-
a^-_{-n} a^+_n+a^{\mu}_{-n}\cdot a^{\mu}_{n}- 
{\tilde a}^+_{-n} {\tilde a}^-_n-{\tilde a}^-_{-n} 
{\tilde a}^+_n+{\tilde a}^{\mu}_{-n}\cdot {\tilde a}^{\mu}_{n})\cr
&+\sum_{r>0}(b^i_{-r}\cdot b^i_{r}+{\tilde b}^i_{-r}\cdot {\tilde b}^i_{r})\cr
&+\sum_{p>0} p(-\psi^+_{-p} \psi^-_p-
\psi^-_{-p} \psi^+_p+\psi^{\mu}_{-p}\cdot \psi^{\mu}_{p}- 
{\tilde \psi}^+_{-p} {\tilde \psi}^-_p-{\tilde \psi}^-_{-p} 
{\tilde \psi}^+_p +{\tilde \psi}^{\mu}_{-p}\cdot {\tilde \psi}^{\mu}_{p})\cr
&+\sum_{p+{1\over 2}>0}(p+{1\over 2})
(\lambda^i_{-p-{1\over 2}}\cdot \lambda^i_{p+{1\over 2}}
+{\tilde \lambda}^i_{-p-{1\over 2}}\cdot 
{\tilde \lambda}^i_{p+{1\over 2}})+a_{\alpha}^{s}]+H_{ghost}\cr}}
where the index $s=1$ denotes the NS-NS sector while $s=2$ indicates the 
RR sector, $a_0^{1}=-1$ and the other $a_i^{s}=0$.

Partition function amplitudes where both pulses are a function of $x^+$
or $x^-$ are supersymmetric, and one obtains the usual result up to
the zero mode contribution--these amplitudes vanish.  When we take pulses
${\vec Y}_1 (x^-)$ and ${\vec Y}_2 (x^+)$, supersymmetry is broken, and
the oscillator mode contributions of 
${\cal O}({\vec Y}_{2+,\eta_2})$ and ${\cal O}({\vec Y}_{1-,\eta_1})$
are relevant.  An exact calculation is perhaps possible, but since we will
be considering low energies compared to the string scale, we will
solve the equations \bc\ to linear order in oscillators  
and to first derivatives of ${\vec Y}$.  One obtains
\eqn\oexpand{\eqalign{{\cal O}({\vec Y}_{+,\eta} )&=O_{osc}\cdot O_0\cr
O_{osc}&=\exp\{ -2\sum_{n>0}{1\over n}\partial_+{\vec Y}_+\cdot
(a_{-n}^+{\tilde{\vec a}}_{-n} +{\vec a}_{-n}{\tilde a}^+_{-n}
-\partial_+{\vec Y}_+
a^+_{-n}{\tilde a}^+_{-n})\cr
&-2i\eta\sum_{p>0}\partial_+{\vec Y}_+\cdot
(\psi^+_{-p}{\tilde{\vec\psi}}_{-p}+{\vec\psi}_{-p}{\tilde\psi}^+_{-p}
-\partial_+{\vec Y}_+\psi^+_{-p}{\tilde\psi}^+_{-p})\}\cr
O_0&=\exp\{ -i{\vec Y}(x^+)\cdot {\vec q}+2i\eta
{\tilde\psi}_0^+{\bar\psi}^2\partial_+ Y_+^{(2)}-2{\tilde\psi}_0^+{\tilde\psi}^2
\partial_+{\bar Y}_+^{(2)}\}\cr}}
where $Y_+^{(2)}={-iY_+^2+Y_+^3\over \sqrt{2}}$.

To calculate the nonzero mode oscillator contribution to the nonsupersymmetric
amplitude, one can go to a T-dual configuration of threebranes at the
orbifold point and then Wick rotate in two directions to obtain 
threebrane instantons and a time direction corresponding to one of the
untwisted transverse directions. (One can alternatively do a direct calculation
on the fermions and use supersymmetry.)  
The Wick rotation makes the electric field
imaginary.  The matrix rotating the oscillators belongs to
$SO(4,{\bf C})$ and takes the form
\eqn\wavematrix{M_+=\pmatrix{1+\partial_+ {\vec Y}_+\cdot\partial_+{\vec Y}_+&
i\partial_+ {\vec Y}_+\cdot\partial_+{\vec Y}_+&\sqrt{2}i\partial_+ Y_+^2&
\sqrt{2}i\partial_+ Y_+^3\cr i\partial_+ {\vec Y}_+\cdot\partial_+{\vec Y}_+&
1-\partial_+ {\vec Y}_+\cdot\partial_+ {\vec Y}_+&-\sqrt{2}\partial_+ Y_+^2&
-\sqrt{2}\partial_+ Y_+^3\cr -\sqrt{2}i\partial_+ Y_+^2&
\sqrt{2}\partial_+ Y_+^2&1&0\cr
-\sqrt{2}i\partial_+ Y_+^3&
\sqrt{2}\partial_+ Y_+^3&0&1\cr}.}
The eigenvalues of this matrix are all equal to one, and there are two
independent eigenvectors.  The action of $M_-$ alone commutes
with the Hamiltonian so that the relevant rotation for the nonsupersymmetric
case is $R=M_{1-}^t M_{2+}$.  Let $y_+=|\partial_+{\vec Y}_{2+}|$,
$y_-=|\partial_-{\vec Y}_{1-}|$, and $\cos\theta={\partial_+{\vec Y}_{2+}\cdot
\partial_-{\vec Y}_{1-}\over y_+ y_-}$.  We find that the eigenvalues
of $R$ take the form $\lambda_1=e^{i\pi\alpha}$, $\lambda_2=e^{-i\pi\alpha}$,
$\lambda_3=e^{i\pi\beta}$, and $\lambda_4=e^{-i\pi\beta}$ where we trust
that the context of using $\alpha$ and $\beta$ will preclude confusion
with the subscripts of boundary states and
\eqn\reigen{\eqalign{&\cos{\pi\alpha}=1-2y_+ y_-\cos\theta +y_+^2 y_-^2\cr
&+\sqrt{(-2y_+ y_-\cos\theta +y_+^2 y_-^2)^2+4y_+^2 y_-^2{\sin^2 \theta}}\cr
&\cos{\pi\beta}=1-2y_+ y_-\cos\theta +y_+^2 y_-^2\cr
&-\sqrt{(-2y_+ y_-\cos\theta +y_+^2 y_-^2)^2 
+4y_+^2 y_-^2{\sin^2 \theta}}.\cr}}
One can see that $\alpha$ is imaginary for $\theta\neq 0$ 
while $\beta$ is imaginary for large $y_+ y_-$.
At least one of the two 
is always imaginary for both $y_+$ and $y_-$ not zero if $\theta\neq 0$.

   The amplitudes are as follows:
\eqn\ampnonsusy{\eqalign{\int_0 ^{\infty} dl 
<F_1 ,{\vec Y}_{1-}|e^{-lH}|F_1 ,{\vec Y}_{2+}>&=
{c_1^2\over 4\pi^2}\int dx^+ dx^-\int_0 ^{\infty} {dl\over l^2}
e^{-b^2 (x^+ ,x^-)\over 2l} Z_{F_1}\cr
\int_0^{\infty} dl 
<F_0 ,{\vec Y}_{1-}|e^{-lH}|F_0 ,{\vec Y}_{2+}>&=
{c_2^2\over 16\pi^4}\int dx^+ dx^-\int_0 ^{\infty} {dl\over l^4}
e^{-b^2 (x^+ ,x^- )\over 2l} Z_{F_0}\cr}}
where $q=e^{-2\pi l}$ and
\eqn\zpadefs{\eqalign{Z_{F_1}&=\prod_i {f_2^4(q)f_3^{\lambda_i}(q)
\over f_4^4(q)f_1^{\lambda_i}(q)}-\prod_i {f_3^4(q)f_2^{\lambda_i}(q)
\over f_4^4(q)f_1^{\lambda_i}(q)}Z_0\cr 
Z_{F_0}&=\prod_i {f_3^4(q)f_3^{\lambda_i}(q)
\over f_1^4(q)f_1^{\lambda_i}(q)}-\prod_i {f_4^4(q)f_4^{\lambda_i}(q)
\over f_1^4(q)f_1^{\lambda_i}(q)}-\prod_i {f_2^4(q)f_2^{\lambda_i}(q)
\over f_1^4(q)f_1^{\lambda_i}(q)}Z_0\cr
Z_0&=1-y_+ y_-\cos\theta\cr}} 
with
\eqn\ffudefs{\eqalign{b(x^+ ,x^-)&=|{\vec Y}_{2+}-{\vec Y}_{1-}|\cr
f_1^{\lambda}(q)&=q^{1\over 12}\prod_{n=1}^{\infty}(1-\lambda q^{2n})\cr
f_2^{\lambda}(q)&=\sqrt{2}q^{1\over 12}
\prod_{n=1}^{\infty}(1+\lambda q^{2n})\cr
f_3^{\lambda}(q)&=q^{-1\over 24}\prod_{n=1}^{\infty}(1+\lambda q^{2n-1})\cr
f_4^{\lambda}(q)&=q^{-1\over 24}\prod_{n=1}^{\infty}(1-\lambda q^{2n-1}).\cr}}
The constants $c_1$ and $c_2$ can be determined from the supersymmetric
case by comparing open and closed cylinders.  We find that
$c_2={\pi\over 2}c_1=\pi^2\alpha' c_1$ where $c_1^2={\pi\over 4}
={\alpha'\pi^2\over 2}$.  The potentials vanish when either $y_+$ or
$y_-$ is zero.

The total amplitude in the field theory ($l\rightarrow\infty$) limit is
\eqn\ampllarge{\eqalign{\lim_{l\rightarrow\infty}
{\cal A}_{\alpha\beta}^{-+}&=
{1\over 8\pi^2\alpha'}(-1+4\delta_{\alpha\beta})\int dx^+ dx^- 
(\int_l ^{\infty} {dl'\over {l'}^2}
e^{-b^2 (x^+ ,x^-)\over 4\pi l'\alpha'}y_+ y_-\cos\theta )\cr
&+{1\over 64\pi^2\alpha'}\int dx^+ dx^- (\int_l ^{\infty} {dl'\over {l'}^4}
e^{-b^2 (x^+ ,x^- )\over 4\pi l'\alpha'} {y_+^2 y_-^2\over 2}).\cr}}
The absolute value of the potential decreases with separation of
the D-strings, but the sign of the potential can change as a function
of the two-dimensional position since $\theta$ depends on
$x_+$ and $x_-$.  The maximum nonunitary eigenvalue $\lambda$ generates a
logarithmic divergence in the amplitude as $\lambda q^2\rightarrow 1$
and should be indicative of a decay process.  For small $l$ a naive modular
transformation continued to nonunitary $\lambda$, $l\rightarrow{1\over 2t}$, 
reveals a tachyon in the
open string spectrum for $b(x^+ ,x^-)\sim\sqrt{2\pi^2\alpha'}$, 
and the amplitude
gains an imaginary piece.  

Although the potential vanishes when the wave is
unidirectional, supersymmetry is not generally restored continuously.
To analyze this issue, we multiply ${\vec Y}_{1-}$ by $0<\epsilon<<1$
and examine the correlation function
\eqn\epscorr{{\cal A}^{-+}_{F_1,\epsilon}=\epsilon
\int_0^{\infty} dl 
<F_1,\eta_1|S^{\dagger}_{-,\eta_1}e^{-lH}|F_1,\eta_2,{\vec Y}_{2+}>.}
The purely bosonic part of this correlation function vanishes by 
supersymmetry.  The fermion zero mode piece in the even spin
structure of the RR sector (the odd spin structure vanishes due to
zero modes) takes the form
\eqn\epscorzerr{{\epsilon\over\alpha'}\int dx^+ dx^-
\int_0^{\infty} {dl\over l^2}{f_2^4(q)f_3^4(q)\over f_1^4(q)f_4^4(q)}
e^{-{b^2(x^+,x^-)\over 4\pi l\alpha'}}(y_+ y_-\cos\theta +O(\alpha'^{2})).}
A modular transformation shows that there is a tachyon for
$b(x^+,x^-)<b_{crit}=\sqrt{2\pi^2\alpha'}$.  The NS-NS terms
do not cancel the tachyon.  There is also a divergence in the
$\alpha'$ expansion for $l\rightarrow 0$.  Without a detailed 
calculation, the expansion for small $l$ is a series with terms of the form
$({2\pi\alpha'\over l})^{2n}$ multiplied by $(2n)^{th}$ derivatives 
of $y_+$ and $y_-$.  For $b>>b_{crit}$, the exponential damping
factor should cancel this divergence.  The divergence for small
$b$ is localized in a region of the D-strings if the pulses have finite 
energy.  Because the divergence exists for infinitesimal $\epsilon$,
supersymmetry is not restored continuously unless $b>>b_{crit}$.
The target space supersymmetry being broken by the presence of
the D-string with an antiparallel travelling wave is chiral,
and one might expect a discontinuity when the mass of the
string communicating this breaking is sufficiently small.

The above potentials imply that the vacuum is not stable unless either
$y_+$ or $y_-$ is zero or, alternatively, $b(x^+ ,x^-)\rightarrow\infty$
for $x^+$ and $x^-$ such that $y_+ y_-\neq 0$.  The interaction is
localized in space and time on the D-strings.  Depending on the 
details of the pulses, the interacting pieces are pushed apart or they 
attract.  In the case of attraction, a tachyon develops at a critical
separation, and one expects the system to decay to a supersymmetric
set of D-strings with a unidirectional travelling wave.  Both
possibilities can occur in different regions of the D-string 
worldvolume and the two possibilities are exchanged by interchanging
the $\alpha=\beta$ and  $\alpha\neq\beta$ boundary states.  For the radiation
calculation of the next section, we will take $b>>b_{crit}$
and $g_s=\lim_{R\rightarrow\infty}e^{\phi}<<1$ so that the 
perturbative calculation is approximately valid.  Also, we will take
only the lowest order in the $\alpha'$ derivative expansion as we 
will be studying radiation in the field theory limit.  Note that $y_+$
and $y_-$ are dimensionless since ${\vec Y}$ has the dimension of length,
and without additional input we would need to include arbitrary
powers of $y_+$ and $y_-$ obtained from the boundary state of
equation \oexpand .  To make the radiation calculation less than horrendous
as well as to minimize back reaction effects, we will take $y_+ ,y_- <<1$.

\newsec{Gravitational Radiation}

Calculation of gravitational radiation from a cosmic string with
travelling waves was performed in \twelve .
In this section we calculate the massless NS-NS closed string emission from
the interaction of two D-strings with antiparallel travelling waves.  A
similar calculation of radiation from moving D-particle interactions is
extensively discussed in \nine , and many of the results there will
be useful in our calculation.  We want to determine the following
amplitude:
\eqn\amprad{{\cal A}^{-+}_{\alpha\beta}(p)=
\int_0^\infty dl \int_0^l d\tau <B_{\alpha},
\vec{Y}_{1-}|e^{-lH}V_p(\sigma ,\tau)|B_{\beta},\vec{Y}_{2+}>}
where we approximate the boundary states as in equation \oexpand\ with
$y_+, y_- <<1$, the closed string vertex operator
\eqn\vertexop{V_p(z,{\bar z})=e_{\mu\nu}(\partial X^{\mu}-{1\over 2}
p\cdot\Psi\Psi^{\mu})({\bar\partial} X^{\nu}+{1\over 2}
p\cdot{\tilde{\Psi}}{\tilde\Psi}^{\nu})e^{ip\cdot X},}
and $2p^+ p^-=\|{\vec p}\|^2\equiv p^2$.  
We restrict the index $\mu\in\{+,-,2,3\}$.  Here, $X^{\mu}=X^{\mu}(z)+
{\tilde X}^{\mu}({\bar z})$.
Because there are
extra fermion zero modes for the ${\bf Z}_2\times {\bf Z}_2$ orbifold,
the odd spin structure vanishes unless the $V_p$ insertion 
has these zero modes.  If we consider polarizations in the extra dimensions,
then there will be a term with two zero modes.  We will not discuss this
case further here.

Our gauge choice is the following.  The axion has antisymmetric
components $e_{+i}={-a^p p^- p_i\over p^2}
-b^p{p^-\epsilon_{ij} p^j\over p^2}$, 
$e_{-i}={a^p p^+ p_i\over p^2}-b^p{p^+\epsilon_{ij} p^j\over p^2}$, 
$e_{+-}=a^p$, and $e_{ij}=b^p\epsilon_{ij}$.
The dilaton has a symmetric polarization tensor with nonvanishing components
$e_{++}={\phi^p(p^-)^2\over p^2}$, $e_{--}={\phi^p(p^+)^2\over p^2}$, 
$e_{+-}={-\phi^p\over 2}$,
and $e_{ij}=\phi^p(\delta_{ij}-{p_i p_j\over p^2})$ for $i,j\in\{2,3\}$.
The graviton has symmetric components $e_{++}=-h_2^p{(p^-)^2\over p^2}$,
$e_{--}=h_2^p{(p^+)^2\over p^2}$,
$e_{i+}=h_1^p{p^-\epsilon_{ij} p^j\over p^2}+h_2^p{p^- p_i\over 2p^2}$, and
$e_{i-}=-h_1^p{p^+\epsilon_{ij} p^j\over p^2}-h_2^p{p^+ p_i\over 2p^2}$ 
with the others vanishing and satisfies 
$\eta^{\mu\nu}e_{\mu\nu}=0$.  All polarization tensors satisfy
$p^{\mu}e_{\mu\nu}=0$, 
and the dilaton has $\eta^{\mu\nu}e_{\mu\nu}=2\phi^p$.
The index $p$ indicates that these components are a Fourier transform that
will be defined more precisely near the finish of this calculation.
This gauge is consistent with working in the rescaled metric \metric .
We will use the notation 
${\vec v_1}\times {\vec v_2}=v_1^i v_2^j\epsilon_{ij}$.
Note that $h_1^p$ is gauge invariant while $h_2^p$ mixes with the
axion so that $h_2^p -2a^p$ is gauge invariant.  As we realized after
doing the calculation, $b^p$ is the fourth physical degree of freedom
that can either be packaged as the $ij$ component of the axion or
as part of the graviton.  

The amplitude can be split into a zero mode and oscillator part.  
Defining $l'=l-\tau$ so that $l'=0$ at the boundary with pulse
${\vec Y}_1(x^-)$ and $l'=l$ at the boundary with ${\vec Y}_2(x^+)$,
the amplitude between two $F_1$ D-strings can be written as
(see \nine\ for details)
\eqn\maineq{\eqalign{{\cal A}_{F_1}(p)&=c_1^2\int dx^+ dx^-
(e^{i{\vec Y}_2(x^+)
\cdot {\vec p}-ip^+x^- -ip^- x^+})\cdot\cr
&(\int_0^{\infty}d\tau
\int_0^{\infty}dl'\int {d^4k\over(2\pi)^4} e^{i{\vec k}\cdot{\vec b}(x^+ ,x^-)}
e^{-l'k^2/2} e^{-\tau q^2/2}<e^{ip\cdot X}>_{osc}{\cal N}_{F_1}).\cr}}
Note that $k^+ =p^+$, 
$q^- =-p^-$, 
and ${\vec k}={\vec q}+{\vec p}$
with $k^- =q^+ =0$.  Also, ${\vec b}(x^+ ,x^-)=
{\vec Y}_1(x^-)-{\vec Y}_2(x^+)$.  To lowest order in derivatives, the 
integrals over $x^+$ and $x^-$ enforce the further conditions that
$p^+=\partial_-{\vec Y}_{1-}\cdot{\vec k}$
and $p^-=-\partial_+{\vec Y}_{2+}\cdot{\vec q}$.
The factor
\eqn\Namp{{\cal N}_{F_1}\equiv\sum_{s,a}Z_{F_1}^{s,a}{\cal M}^{s,a}
\epsilon^{s,a}_{F_1}}
where 
\eqn\zpart{Z_{F_1}^{s,a}
=<F_1,\eta_1,\vec{Y}_{1-}|e^{-lH}|F_1,\eta_2,\vec{Y}_{2+}>^{s,a}_{osc},}
$a=\pm 1$
with $\eta_1\eta_2=a$, and $\epsilon^{s,a}_{F_1}=\pm 1$ is determined by
\fullbo .  The $Z_{F_1}^{s,a}$ can be read from equation \zpadefs\ where
we only take linear terms in $y_+$ and $y_-$.

In the even spin structure correlation functions are defined as
\eqn\defcorreven{<{\cal O}(\sigma ,\tau)>^s_{osc}\equiv
{<F_i ,\eta_1,\vec{Y}_{1-}|
e^{-lH}{\cal O}(\sigma ,\tau)|F_i ,\eta_2,\vec{Y}_{2+}>^s_{osc}
\over Z^s_{F_i}}.}
Bosonic correlators can be calculated with $y_+=y_-=0$ since they get
multiplied by a fermion zero mode term proportional to $y_+ y_-$.
The correlation function

\eqn\expp{<e^{ip\cdot X}>_{osc}=\prod_{n=0}^{\infty}
(1-q^{2n}e^{-4\pi\tau})^{-{p^2\over 2\pi}}
(1-q^{2n}e^{-4\pi l'})^{-{p^2\over 2\pi}}.}
The odd spin structure will only contribute to the $F_0$ amplitudes in
the NS-NS sector and is defined as above.

The factor ${\cal M}^{sa}$ takes the form
\eqn\mfactor{\eqalign{{\cal M}^{sa}&=e_{\mu\nu}\{ <\partial X^{\mu}
{\bar\partial}X^{\nu}>_{osc}-<\partial X^{\mu} p\cdot X>_{osc}
<{\bar\partial}X^{\nu}p\cdot X>_{osc}\cr
&+{1\over 4}(<p\cdot\Psi p\cdot{\tilde\Psi}>^{sa}
<\Psi^{\mu}{\tilde\Psi}^{\nu}>^{sa}\cr &-<p\cdot\Psi\Psi^{\mu}>^{sa}
<p\cdot{\tilde\Psi}{\tilde\Psi}^{\nu}>^{sa}+<p\cdot{\tilde\Psi}\Psi^{\mu}>^{sa}
<p\cdot\Psi{\tilde\Psi}^{\nu}>^{sa})\cr
&+{i\over 2}(<\partial X^{\mu} p\cdot X>_{osc}
<p\cdot{\tilde\Psi}{\tilde\Psi}^{\nu}>^{sa}-
<{\bar\partial}X^{\nu}p\cdot X>_{osc}<p\cdot\Psi\Psi^{\mu}>^{sa})\cr
&-{1\over 2}k^{\mu}(i<{\bar\partial}X^{\nu}p\cdot X>_{osc}
+{1\over 2}<p\cdot{\tilde\Psi}{\tilde\Psi}^{\nu}>^{sa})\cr
&+{1\over 2}k^{\nu}(i<\partial X^{\mu} p\cdot X>_{osc}
-{1\over 2}<p\cdot\Psi\Psi^{\mu}>^{sa})-{1\over 4}k^{\mu}k^{\nu}\}.\cr}}
The nonzero correlators, where we need to keep track of $\eta$, 
are the following:
\eqn\corrone{<\partial X^+(z){\tilde X}^-({\bar z})>_{osc}=
-<{\bar\partial}{\tilde X}^+({\bar z})X^-(z)>_{osc}={i\over 2} K(\tau,l)}
\eqn\corrtwo{<\partial X^i(z){\tilde X}^j({\bar z})>_{osc}=
\delta^{ij}{i\over 2} K(\tau,l)} 
\eqn\corrthree{\eqalign{&<\Psi^+(z){\tilde\Psi}^-({\bar z})>^{sa}=
<\Psi^-(z){\tilde\Psi}^+({\bar z})>^{sa}=\cr
&{-i\eta_1\over 2 Z_0}\delta^s_2\delta^a_+ +i\eta_1 F^{sa}
-4i\eta_1{\vec y}_+ 
\cdot{\vec y}_-(H^{sa}(l',l)+\eta_1\eta_2 H^{sa}(\tau ,l))\cr}}
\eqn\corrfour{\eqalign{<\Psi^i(z){\tilde\Psi}^j({\bar z})>^{sa}
&=({-i\eta_1\over
2Z_0}\delta^{ij}-{i\eta_1\over
2Z_0}(y_+^i y_-^j +y_+^j y_-^i))\delta^s_2\delta^a_+\cr
&+i\eta_1 \delta^{ij}F^{sa}-4i\eta_1(y_+^i y_-^j +y_+^j y_-^i)
(H^{sa}(l',l)+\eta_1\eta_2 H^{sa}(\tau ,l))\cr}}
\eqn\corriplus{<\Psi^i(z){\tilde\Psi}^+({\bar z})>^{sa}
={-i\eta_1\over 2 Z_0}y^i_-\delta^s_2\delta^a_+-2i\eta_2 y^i_- G^{sa}(l',l)}
\eqn\corriminus{<\Psi^i(z){\tilde\Psi}^-({\bar z})>^{sa}
={-i\eta_1\over 2 Z_0}y^i_+\delta^s_2\delta^a_+ 
-2i\eta_1 y^i_+ G^{sa}(\tau ,l)}
\eqn\corrplusplus{<\Psi^+(z){\tilde\Psi}^+({\bar z})>^{sa}=
-2i\eta_2{\vec y}_-\cdot{\vec y}_- G^{sa}(l',l)}
\eqn\corrminmin{<\Psi^-(z){\tilde\Psi}^-({\bar z})>^{sa}=
-2i\eta_1{\vec y}_+\cdot{\vec y}_+ G^{sa}(\tau ,l)}
\eqn\corrfive{\eqalign{&<\Psi^+(z)\Psi^-(z)>^{sa}=-N_{\infty}^{osc}\cr
&<{\tilde\Psi}^+({\bar z}){\tilde\Psi}^-({\bar z})>^{sa}=
{1-2Z_0\over Z_0}\delta^s_2\delta^a_+ -N_{\infty}^{osc}\cr}}
\eqn\corrsix{\eqalign{&<\Psi^i(z)\Psi^j(z)>^{sa}=
\delta^{ij}({1\over 2}\delta^s_2\delta^a_+
+N_{\infty}^{osc})\cr
&<{\tilde\Psi}^i({\bar z}){\tilde\Psi}^j({\bar z})>^{sa}=
({1\over 2}\delta^{ij}+{i\over 2}\epsilon^{ij}{Z_0-1\over Z_0})
\delta^s_2\delta^a_+ +\delta^{ij}N_{\infty}^{osc}\cr}}
\eqn\corrseven{<\Psi^i(z)\Psi^+(z)>^{sa}=
<{\tilde\Psi}^i({\bar z}){\tilde\Psi}^+({\bar z})>^{sa}=
{-y_-^i\over 2 Z_0}\delta^s_2\delta^a_+}
\eqn\correight{<\Psi^i (z)\Psi^-(z)>^{sa}=
<{\tilde\Psi}^i({\bar z}){\tilde\Psi}^-({\bar z})>^{sa}=
{y_+^i\over 2 Z_0}\delta^s_2\delta^a_+}
\eqn\defone{K(\tau , l)=-\sum_{n=0}^{\infty}({q^{2n}e^{-4\pi\tau}\over
1-q^{2n}e^{-4\pi\tau}}-{q^{2n}e^{-4\pi l'}\over
1-q^{2n}e^{-4\pi l'}})}
\eqn\deftwo{\eqalign{&F^{2+}
=-\sum_{n=0}^{\infty}(-1)^{n}({q^{2n}e^{-4\pi\tau}\over
1-q^{2n}e^{-4\pi\tau}}+{q^{2n}e^{-4\pi l'}\over
1-q^{2n}e^{-4\pi l'}})\cr
&F^{1\pm}=-\sum_{n=0}^{\infty}(\mp )^{n}({q^{n}e^{-2\pi\tau}\over
1-q^{2n}e^{-4\pi\tau}}\pm{q^{n}e^{-2\pi l'}\over
1-q^{2n}e^{-4\pi l'}})\cr}}
\eqn\defthree{\eqalign{G^{2+}(x,l)&=\sum_{n=0}^{\infty}(-1)^n (n+1){q^{2n}
e^{-4\pi x}\over 1-q^{2n}e^{-4\pi x}}\cr
G^{1\pm }(x,l)&=\sum_{n=0}^{\infty}(\mp )^n (n+1){q^{n}
e^{-2\pi x}\over 1-q^{2n}e^{-4\pi x}}\cr}}
\eqn\deffour{\eqalign{H^{2+}(x,l)&=\sum_{n=0}^{\infty}(-1)^n {(n+2)!
\over n! 2!}{q^{2n+2}
e^{-4\pi x}\over 1-q^{2n+2}e^{-4\pi x}}\cr
H^{1\pm }(x,l)&=\sum_{n=0}^{\infty}(\mp )^n {(n+2)!
\over n! 2!}{q^{n+1}
e^{-2\pi x}\over 1-q^{2n+2}e^{-4\pi x}}.\cr}}
To simplify notation ${\vec y}_+ =\partial_+{\vec Y}_{2+}$ and
${\vec y}_- =\partial_-{\vec Y}_{1-}$.  
The infinite constant $N_{\infty}^{osc}$
does not contribute to  ${\cal M}^{sa}$ because $p^{\mu}e_{\mu\nu}=0$.
Note that $H^{2+}$ will not contribute in the field theory limit.

The axion emission comes from the asymmetry of left and right zero mode
fermions in the even spin RR sector.  We obtain the term
\eqn\axi{{\cal M}^{sa}_{ax}={a^p\over 4}({1-2 Z_0\over Z_0})(p^2-{\vec k}
\cdot {\vec p})\delta^s_2\delta^a_+ .}
There is no dependence on nonzero mode fermion correlations.
For the dilaton and the graviton the term ${\cal M}^{sa}$ simplifies to
\eqn\dil{\eqalign{{\cal M}^{sa}_{\phi}&=\phi^p\{ ({p^2\over 16}
+{1\over 8}({\vec p}\cdot{\vec y}_+) ({\vec p}\cdot{\vec y}_-)
-{1\over 8} p^2 {\vec y}_+\cdot{\vec y}_-\cr
&+{1\over 8}({\vec p}\cdot{\vec y}_+)({\vec k}\cdot{\vec y}_-)
-{1\over 8}({\vec q}\cdot{\vec y}_+)({\vec p}\cdot{\vec y}_-)
-{1\over 4}({\vec k}\cdot{\vec y}_+)({\vec k}\cdot{\vec y}_-)
-{1\over 4}({\vec q}\cdot{\vec y}_+)({\vec k}\cdot{\vec y}_-)\cr
&+{1\over 4}{{\vec k}\cdot{\vec p}\over p^2}
({\vec q}\cdot{\vec y}_+)({\vec p}\cdot{\vec y}_-)
+{1\over 4}{{\vec k}\cdot{\vec p}\over p^2}
({\vec p}\cdot{\vec y}_+)({\vec k}\cdot{\vec y}_-)
-{i\over 8}({\vec p}\times{\vec k})({\vec y}_+\cdot{\vec y}_-) )
\delta^s_2\delta^a_+\cr
&+({1\over 2}( p^2 ({\vec y}_+\cdot{\vec y}_- )  
-({\vec p}\cdot{\vec y}_+) ({\vec p}\cdot{\vec y}_- ) )F^{sa}\cr
&+{p^2\over 4}{\vec y}_+\cdot{\vec y}_- 
(G^{sa}(l',l)+G^{sa}(\tau ,l))
-(p^2 {\vec y}_+\cdot{\vec y}_- -2({\vec p}\cdot{\vec y}_+) 
({\vec p}\cdot{\vec y}_-))
(H^{sa}(l',l)+H^{sa}(\tau ,l)))\delta^s_2\delta^a_+\cr
&-a({\vec p}\cdot{\vec y}_+) ({\vec p}\cdot{\vec y}_-) 
G^{sa}(l',l)G^{sa}(\tau ,l)\cr
&+2(p^2 {\vec y}_+\cdot{\vec y}_- -2({\vec p}\cdot{\vec y}_+) 
({\vec p}\cdot{\vec y}_-))F^{sa}(H^{sa}(l',l)+aH^{sa}(\tau ,l))
-{k^2\over 4}+{1\over 4}{({\vec k}\cdot{\vec p})^2\over p^2}
-{1\over 16}p^2 \}\cr}}
\eqn\gravhone{\eqalign{{\cal M}^{sa}_{h1}&=
h_1^p\{ [{1\over 8}(({\vec y}_+\times{\vec p})( {\vec y}_-\cdot{\vec p})
-({\vec y}_+\cdot{\vec p})({\vec y}_-\times{\vec p}))
-i{p^2\over 8}({\vec y}_+\cdot{\vec y}_-) -{1\over 4}({\vec k}\times 
{\vec p}) ({\vec y}_+\cdot{\vec y}_-)\cr
&+{1\over 8}({\vec y}_+\cdot{\vec q})({\vec y}_-\times{\vec p})
+{1\over 4}({\vec y}_+\times{\vec p})( {\vec y}_-\cdot{\vec k})\cr 
&-{1\over 4}{({\vec k}\times{\vec p})\over p^2}
({\vec q}\cdot{\vec y}_+)({\vec p}\cdot{\vec y}_-)
+{1\over 4}{({\vec k}\times{\vec p})\over p^2}
({\vec p}\cdot{\vec y}_+)({\vec k}\cdot{\vec y}_-)]\delta^s_2\delta^a_+\cr
&+[{1\over 4}({\vec y}_+\cdot{\vec p})({\vec y}_-\times{\vec p}) 
(G^{sa}(\tau ,l)-G^{sa}(l',l))
+{1\over 4}({\vec y}_+\times{\vec p})({\vec y}_-\cdot{\vec p})
(G^{sa}(\tau ,l)-G^{sa}(l',l))\cr
&-{1\over 2}(({\vec y}_+\cdot{\vec q})({\vec y}_-\times{\vec p}))
(F^{sa}-2G^{sa}(l',l))
-{1\over 2}({\vec y}_+\times{\vec p})({\vec y}_-\cdot{\vec k})
(F^{sa}-2G^{sa}(\tau ,l))]\delta^s_2\delta^a_+\cr
&-a(({\vec y}_+\times{\vec p})( {\vec y}_-\cdot{\vec p})
-({\vec y}_+\cdot{\vec p})({\vec y}_-\times{\vec p}))G^{sa}(\tau ,l)
G^{sa}(l',l)\cr
&-2a({\vec y}_+\cdot{\vec q})({\vec y}_-\times{\vec p})G^{sa}(l',l)F^{sa}
-2({\vec y}_+\times{\vec p})({\vec k}\cdot{\vec y}_-)G^{sa}(\tau ,l)F^{sa}
-{1\over 4}({\vec k}\times{\vec p})K-{1\over 4}({\vec k}\times{\vec p})\}\cr}}
\eqn\gravhtwo{\eqalign{{\cal M}^{sa}_{h2}
&=h_2^p\{[{p^2\over 16}({\vec y}_+\cdot{\vec y}_-)
-{1\over 8}({\vec k}\cdot {\vec p})({\vec y}_+\cdot{\vec y}_-)
+{3\over 16}({\vec q}\cdot{\vec y}_+)({\vec p}\cdot{\vec y}_-)
+{1\over 8}({\vec p}\cdot{\vec y}_+)({\vec k}\cdot{\vec y}_-)\cr
&-{1\over 8}{({\vec k}\cdot {\vec p})\over p^2}
({\vec q}\cdot{\vec y}_+)({\vec p}\cdot{\vec y}_-)
+{1\over 8}{({\vec k}\cdot {\vec p})\over p^2}
({\vec p}\cdot{\vec y}_+)({\vec k}\cdot{\vec y}_-)]\delta^s_2\delta^a_+\cr
&+[{1\over 4}({\vec p}\cdot{\vec y}_+)( {\vec p}\cdot{\vec y}_-)
(G^{sa}(\tau ,l)-G^{sa}(l',l))\cr
&+{1\over 2}({\vec q}\cdot{\vec y}_+)({\vec p}\cdot{\vec y}_-)
(G^{sa}(l',l)-F^{sa})
+{1\over 2}({\vec p}\cdot{\vec y}_+)({\vec k}\cdot{\vec y}_-)
(G^{sa}(\tau ,l)-F^{sa})]\delta^s_2\delta^a_+\cr
&-{1\over 8}p^2 K+{1\over 16}p^2-{1\over 8}({\vec k}\cdot {\vec p})\}.\cr}} 

The field theory ($l\rightarrow\infty$) limits of $K$, $F^{sa}$,
and $G^{sa}$ are
\eqn\limits{\eqalign{K&\rightarrow -(f(\tau)-f(l'))\cr
F^{2+}&\rightarrow -(f(\tau)+f(l'))\cr
G^{2+}(x,l)&\rightarrow f(x)\cr
G^{1\pm}(\tau ,l)G^{1\pm}(l',l)&\rightarrow(-1+f(\tau)+f(l'))q\cr}}
where $f(x)={e^{-4\pi x}\over 1+e^{-4\pi x}}$.  Integration by parts
has been used to calculate ${\cal M}^{sa}$ with
\eqn\intbyparts{\int_0^{\infty}d\tau\int_0^{\infty}dl'({\partial}_{\tau}
-{\partial}_{l'})\{ e^{-{q^2+m^2\over 2}\tau}e^{-{k^2+m^2\over 2}l'}
<e^{ip\cdot X}>_{osc}\}=0}
where the infinite surface terms have been set to zero 
by an analytic continuation from $p^2<0$.
This integration by parts in the $l\rightarrow\infty$ limit allows
one to replace $f(\tau)$ by ${-1\over 4}{q^2+m^2\over p^2}$ and 
$f(l')$ by ${-1\over 4}{k^2+m^2\over p^2}$ where $m^2$ is the momentum in the
zero mode directions orthogonal to the four Minkowski dimensions.  There
are cancellations between different sectors so that only terms at the
most linear in
the $f(x)$ appear.  
We also make use of the
relation
\eqn\rel{\int_0^{\infty}dx e^{-xa^2\over 2}(1-e^{-4\pi x})^{-b^2\over 2\pi}
={1\over 4\pi}B({a^2\over 8\pi} ,1-{b^2\over 2\pi})
={1\over 4\pi}{\Gamma({a^2\over 8\pi})\Gamma(1-{b^2\over 2\pi})\over
\Gamma({a^2\over 8\pi}+1-{b^2\over 2\pi})}.}
Equation \Namp\ yields 
\eqn\Namptwo{\eqalign{{\cal N}_{F_1}&=\sum_{\eta_1}(Z_{F_1}^{1+}
{\cal M}^{1+}+Z_{F_1}^{2+}{\cal M}^{2+})\cr
{\cal N}_{F_0}&=\sum_{\eta_1}(Z_{F_0}^{1+}
{\cal M}^{1+}-Z_{F_0}^{1-}{\cal M}^{1-}+Z_{F_0}^{2+}{\cal M}^{2+}).\cr}}
Calculating these terms in the $l\rightarrow\infty$ limit yields
\eqn\namplimoa{\lim_{l\rightarrow\infty}{\cal N}^{ax}_{F_1}=
a^p(1-2{\vec y}_+\cdot{\vec y}_-)(p^2-{\vec k}\cdot{\vec p})}
\eqn\namplimza{\lim_{l\rightarrow\infty}{\cal N}^{ax}_{F_0}=
4{\cal N}^{ax}_{F_1}}
\eqn\namplimod{\eqalign{\lim_{l\rightarrow\infty}{\cal N}^{\phi}_{F_1}&=\phi^p
\{ -{1\over 4}p^2+({1\over 4}p^2-{3\over 2}k^2+{1\over 2}{\vec k}\cdot {\vec p}
+{({\vec k}\cdot {\vec p})^2\over p^2}+{i\over 2}{\vec p}\times{\vec k}
-{1\over 2}m_1^2){\vec y}_+\cdot{\vec y}_-\cr
&+{1\over 2}({2k^2-p^2+m_1^2\over p^2})({\vec p}\cdot{\vec y}_+)
({\vec p}\cdot{\vec y}_-)
+2({\vec k}\cdot{\vec y}_+)({\vec k}\cdot{\vec y}_-)\cr
&-({3\over 2}+ {{\vec k}\cdot {\vec p}\over p^2})({\vec p}\cdot{\vec y}_+)
({\vec k}\cdot{\vec y}_-)+({1\over 2}-{{\vec k}\cdot {\vec p}\over p^2})
({\vec k}\cdot{\vec y}_+)({\vec p}\cdot{\vec y}_-)\}\cr}}
\eqn\namplimzd{\eqalign{\lim_{l\rightarrow\infty}{\cal N}^{\phi}_{F_0}&=\phi^p
\{ -p^2+(2p^2+2i{\vec p}\times{\vec k}){\vec y}_+\cdot{\vec y}_-\cr
&-({p^2-6k^2-6{\vec k}\cdot {\vec p}-6m_0^2\over p^2})
({\vec p}\cdot{\vec y}_+)({\vec p}\cdot{\vec y}_-)+8({\vec k}\cdot{\vec y}_+)
({\vec k}\cdot{\vec y}_-)\cr &-(6+4{{\vec k}\cdot {\vec p}\over p^2}) 
({\vec p}\cdot{\vec y}_+)({\vec k}\cdot{\vec y}_-)
+(2-4{{\vec k}\cdot {\vec p}\over p^2})
({\vec k}\cdot{\vec y}_+)({\vec p}\cdot{\vec y}_-)\}\cr}}
\eqn\namplimohone{\eqalign{\lim_{l\rightarrow\infty}{\cal N}^{h_1}_{F_1}&=h_1^p
\{ -{p^2+2{\vec k}\cdot{\vec p}\over 4p^2}({\vec y}_+\times{\vec p})
({\vec y}_-\cdot{\vec p})+
{3p^2+2{\vec k}\cdot{\vec p}\over 4p^2}({\vec y}_+\cdot{\vec p})
({\vec y}_-\times {\vec p})\cr
&-{{\vec k}\cdot{\vec p}\over p^2}
({\vec y}_+\cdot{\vec k})({\vec y}_-\times {\vec p})
-{3p^2-2{\vec k}\cdot{\vec p}\over 2p^2}
({\vec y}_+\times{\vec p})({\vec y}_-\cdot {\vec k})\cr
&+({-p^2+2{\vec k}\cdot{\vec p}\over 4p^2}{\vec k}\times{\vec p}
+{i\over 2}p^2){\vec y}_+\cdot{\vec y}_-\cr
&+{{\vec k}\times{\vec p}\over p^2}(({\vec y}_+\cdot{\vec k})
({\vec y}_-\cdot{\vec p})-({\vec y}_+\cdot{\vec p})({\vec y}_-\cdot{\vec p})
-({\vec y}_+\cdot{\vec p})({\vec y}_-\cdot{\vec k}))\}\cr}}
\eqn\namplimzhone{\eqalign{\lim_{l\rightarrow\infty}{\cal N}^{h_1}_{F_0}&=h_1^p
\{ ({3p^2+k^2-6{\vec k}\cdot{\vec p}+m_0^2\over 2p^2})
({\vec y}_+\times{\vec p})({\vec y}_-\cdot{\vec p})\cr&+
({11p^2+2k^2+2{\vec k}\cdot{\vec p}+2m_0^2\over 2p^2})({\vec y}_+\cdot{\vec p})
({\vec y}_-\times {\vec p})\cr
&-({11p^2+2k^2+2{\vec k}\cdot{\vec p}+2m_0^2\over 2p^2})
({\vec y}_+\cdot{\vec k})({\vec y}_-\times {\vec p})\cr
&-({11p^2+2k^2-6{\vec k}\cdot{\vec p}+2m_0^2\over p^2})
({\vec y}_+\times{\vec p})({\vec y}_-\cdot {\vec k})
+(2ip^2+4{\vec k}\times{\vec p}){\vec y}_+\cdot{\vec y}_-\cr
&+4{{\vec k}\times{\vec p}\over p^2}(({\vec y}_+\cdot{\vec k})
({\vec y}_-\cdot{\vec p})-({\vec y}_+\cdot{\vec p})({\vec y}_-\cdot{\vec p})
-({\vec y}_+\cdot{\vec p})({\vec y}_-\cdot{\vec k}))\}\cr}}
\eqn\namplimohtwo{\eqalign{\lim_{l\rightarrow\infty}{\cal N}^{h_2}_{F_1}&=h_2^p
\{ -{1\over 8}(p^2-2{\vec k}\cdot{\vec p}){\vec y}_+\cdot{\vec y}_-
-({3k^2-{\vec k}\cdot{\vec p}+3m_1^2\over p^2})
({\vec p}\cdot{\vec y}_+)({\vec p}\cdot{\vec y}_-)\cr
&-({p^2-6k^2+2{\vec k}\cdot{\vec p}-6m_1^2\over 4p^2})({\vec k}\cdot{\vec y}_+)
({\vec p}\cdot{\vec y}_-)\cr
&-{3\over 2}({p^2+k^2-{\vec k}\cdot{\vec p}+m_1^2\over p^2})
({\vec p}\cdot{\vec y}_+)
({\vec k}\cdot{\vec y}_-))\}\cr}}
\eqn\namplimzhtwo{\lim_{l\rightarrow\infty}{\cal N}^{h_2}_{F_0}=
4{\cal N}^{h_2}_{F_1}+({1\over 2}p^2-{\vec k}\cdot{\vec p})
{\vec y}_+\cdot{\vec y}_- .}
We are considering radiated energies well below the string scale
so we only take the lowest order terms in $\alpha' p^2$ and utilize 
the relation
\eqn\betalim{\lim_{\alpha' p^2\rightarrow 0}B({x^2\over 8\pi},1-\alpha' p^2)
={8\pi\over x^2}.}
Finally, the amplitudes reduce to 
\eqn\ampgen{\eqalign{\lim_{\alpha' p^2<<1 \atop l\rightarrow\infty}
&{\cal A}_{\alpha\beta}(p)=(\int dx^+ dx^- (e^{i{\vec Y}_2(x^+)
\cdot {\vec p}-ip^+x^- -ip^- x^+})\cdot\cr
&4c_1^2\sum_i\epsilon^i_{\alpha\beta}\int^{\Lambda}{d^2 m_i\over4\pi^2}
\int^{\Lambda} {d^2k\over(2\pi)^2} e^{i{\vec k}\cdot{\vec b}(x^+ ,x^-)}
{{\cal N}_{F_i}\over (k^2+m_i^2)(q^2+m_i^2)})+{\cal A}_0(p)\cr}}
\eqn\ampgenz{\eqalign{\lim_{\alpha' p^2<<1 \atop l\rightarrow\infty}
&{\cal A}_0(p)=\int dx^+ dx^- (e^{i{\vec Y}_2(x^+)
\cdot {\vec p}-ip^+x^- -ip^- x^+})\cdot\cr
&\pi^2 c_1^2\int^{\Lambda}{d^6 m_0\over (2\pi)^6}
\int^{\Lambda} {d^2k\over(2\pi)^2} e^{i{\vec k}\cdot{\vec b}(x^+ ,x^-)}
{{\cal N}_{F_0}\over (k^2+m_0^2)(q^2+m_0^2)}\cr}}
where $\epsilon^i_{\alpha\beta}=\pm 1$ and can be determined from
\frbrane .  We have introduced a cutoff on momentum, 
$\Lambda\sim{1\over\sqrt{\alpha'}}$ because the large momentum region 
picks out the small $l$ region where the above calculation is no 
longer valid, but for $b>>b_{crit}$ taking $\Lambda\rightarrow\infty$
should be okay.  Again, we can do a modular transformation 
$l\rightarrow {1\over 2t}$ to show that there is no divergence so long 
as the average separation of the D-strings is sufficiently large.
We can calculate the above integrals using the formula
\eqn\modbess{{1\over 4\pi^2}\int d^2k {e^{i{\vec k}\cdot{\vec b}}\over
k^2+m^2}={-1\over 2\pi}K_0(mb)}
where $K_0$ is the modified Bessel function of order zero.
We also need the following integrals which can be found in
\thirteen .
\eqn\intone{\eqalign{&\int_0^{\infty} x K_0(ax)dx={\pi\over a^2}\cr
&\int_0^{\infty} x K_0(ax) K_0(bx)dx={\ln{a\over b}\over a^2-b^2}\cr}}
Defining
\eqn\gfunction{g({\vec b},{\vec p})={1\over 4\pi^2}\int d^2x
{\ln{|{\vec b}-{\vec x}|\over |{\vec x}|}e^{i{\vec p}\cdot{\vec x}}
\over b^2-2{\vec b}\cdot{\vec x}}}
where $g({\vec b},{\vec p})$ is finite for $pb>0$, we obtain
\eqn\ampaxtwo{\eqalign
{\lim_{\alpha' p^2<<1 \atop l\rightarrow\infty}
{\cal A}_{\alpha\beta}^{ax}(p)&=4c_1^2(-1+4\delta_{\alpha\beta})
\int dx^+ dx^- (e^{i{\vec Y}_2(x^+)
\cdot {\vec p}-ip^+x^- -ip^- x^+})\cdot\cr
&[a^p(1-2{\vec y}_+\cdot{\vec y}_-)(p^2 g+i{\vec g}_b\cdot{\vec p})]
+{\cal A}_0^{ax}(p)\cr}}
\eqn\ampdiltwo{\eqalign{\lim_{\alpha' p^2<<1 \atop l\rightarrow\infty}
&{\cal A}_{\alpha\beta}^{\phi}(p)=4c_1^2(-1+4\delta_{\alpha\beta})
\int dx^+ dx^- (e^{i{\vec Y}_2(x^+)
\cdot {\vec p}-ip^+x^- -ip^- x^+})\cdot\cr
&[\phi^p\{-{p^2\over 4}g+({1\over 4}p^2 g+\bigtriangleup_b g
-{i\over 2}{\vec g}_b\cdot{\vec p}\cr 
&+{e^{i{\vec p}\cdot{\vec b}}\over 8\pi b^2}
-{{\vec p}\cdot g_{bb}\cdot{\vec p}\over p^2}
+{1\over 2}{\vec p}\times{\vec g}_b)({\vec y}_+\cdot{\vec y}_-)\cr 
&-({e^{i{\vec p}\cdot{\vec b}}\over 8\pi b^2 p^2}
+{1\over 2p^2}\bigtriangleup_b g+{1\over 2}g)
({\vec p}\cdot{\vec y}_+)({\vec p}\cdot{\vec y}_-)\cr
&-2{\vec y}_+\cdot g_{bb}\cdot{\vec y}_- +{3i\over 2}
({\vec p}\cdot{\vec y}_+)({\vec g}_b\cdot{\vec y}_-)
+{1\over p^2}({\vec p}\cdot{\vec y}_+)({\vec p}\cdot g_{bb}\cdot{\vec y}_-)\cr
&-{i\over 2}({\vec g}_b\cdot{\vec y}_+)({\vec p}\cdot{\vec y}_-)
+{1\over p^2}({\vec p}\cdot g_{bb}\cdot{\vec y}_+)
({\vec p}\cdot{\vec y}_-)\}]+{\cal A}_0^{\phi}(p)\cr}}
\eqn\ampgravhonetwo{\eqalign{\lim_{\alpha' p^2<<1 \atop l\rightarrow\infty}
&{\cal A}_{\alpha\beta}^{h_1}(p)=4c_1^2(-1+4\delta_{\alpha\beta})
\int dx^+ dx^- (e^{i{\vec Y}_2(x^+)
\cdot {\vec p}-ip^+x^- -ip^- x^+})\cdot\cr
&[h_1^p\{({i\over 2}p^2 g-{i\over 4}{\vec g}_b\times{\vec p}
-{1\over 2p^2}{\vec p}\cdot g_{bb}\times{\vec p}){\vec y}_+\cdot{\vec y}_-\cr
&+i{{\vec g}_b\times{\vec p}\over p^2}({\vec y}_+\cdot{\vec p})
({\vec y}_-\cdot{\vec p})-{1\over p^2}
({\vec y}_+\cdot g_{bb}\times{\vec p})({\vec y}_-\cdot{\vec p})\cr
&+{1\over p^2}({\vec y}_+\cdot{\vec p})({\vec y}_-\cdot g_{bb}\times{\vec p})
-({1\over 4}g-{i\over 2}{{\vec g}_b\cdot{\vec p}\over p^2})
({\vec y}_+\times{\vec p})({\vec y}_-\cdot{\vec p})\cr
&+({3\over 4}g-{i\over 2}{{\vec g}_b\cdot{\vec p}\over p^2})
({\vec y}_+\cdot{\vec p})({\vec y}_-\times{\vec p})
+{1\over p^2}({\vec y}_+\cdot g_{bb}\cdot{\vec p})({\vec y}_-\times{\vec p})\cr
&+{3i\over 2}({\vec y}_+\times{\vec p})({\vec y}_-\cdot{\vec g}_b)
-{1\over p^2}({\vec y}_+\times{\vec p})
({\vec y}_-\cdot g_{bb}\cdot{\vec p})\}]+{\cal A}_0^{h_1}(p)\cr}}
\eqn\ampgravhtwotwo{\eqalign{\lim_{\alpha' p^2<<1 \atop l\rightarrow\infty}
&{\cal A}_{\alpha\beta}^{h_2}(p)=4c_1^2(-1+4\delta_{\alpha\beta})
\int dx^+ dx^- (e^{i{\vec Y}_2(x^+)
\cdot {\vec p}-ip^+x^- -ip^- x^+})\cdot\cr
&[h_2^p\{ -({1\over 8}p^2 g+{i\over 4}{\vec g}_b\cdot{\vec p})
{\vec y}_+\cdot{\vec y}_-
+({3e^{i{\vec p}\cdot{\vec b}}\over 4\pi b^2p^2 }
-{i\over p^2}{\vec g}_b\cdot{\vec p})
({\vec p}\cdot{\vec y}_+)({\vec p}\cdot{\vec y}_-)\cr
&-{3ie^{i{\vec p}\cdot{\vec b}}\over 4\pi b^4 p^2}({\vec b}\cdot{\vec y}_+)
({\vec p}\cdot{\vec y}_-)
+{i\over 4}({\vec g}_b\cdot{\vec y}_+) ({\vec p}\cdot{\vec y}_-)
+{1\over 2p^2}({\vec y}_+\cdot g_{bb}\cdot{\vec p})({\vec p}\cdot{\vec y}_-)\cr
&+{3ie^{i{\vec p}\cdot{\vec b}}\over 4\pi b^4 p^2}({\vec p}\cdot{\vec y}_+)
({\vec b}\cdot{\vec y}_-)
+{3i\over 2}({\vec p}\cdot{\vec y}_+)({\vec g}_b\cdot{\vec y}_-)
-{3\over 2p^2}({\vec p}\cdot{\vec y}_+)
({\vec y}_-\cdot g_{bb}\cdot{\vec p})\}]+{\cal A}_0^{h_2}(p)\cr}}
where $\bigtriangleup_b$ is the two-dimensional 
Laplacian with respect to ${\vec b}$ while $(g_b)_i=
\partial_{b^i} g({\vec b},{\vec p})$
and $(g_{bb})_{ij}=\partial_{b^i}\partial_{b^j}g({\vec b},{\vec p})$ .  
We leave as an exercise the calculation
of ${\cal A}_0(p)$ which should vanish more
quickly for large $b$.

The function $g$ can be calculated with formulas from \thirteen ,
and we obtain
\eqn\gdetail{\eqalign{g({\vec b},{\vec p})&=
-{e^{{ib p^{(2)}\over 2}}\over 4\pi{\vec p}\times{\vec b}}
\{ Ei({ib p^{(2)}\over 2})
\sinh{{\vec p}\times{\vec b}}\cr
&+Ei({-ib {\bar p}^{(2)}\over 2})e^{{\vec p}\times{\vec b}\over 2}
\sinh{{\vec p}\times{\vec b}\over 2}
-Ei({-ib p^{(2)}\over 2})e^{-{{\vec p}\times{\vec b}\over 2}}
\sinh{{\vec p}\times{\vec b}\over 2}\}}}
where $p^{(2)}={{\vec p}\cdot{\vec b}+i{\vec p}\times{\vec b}\over b}$, 
and the exponential
integral takes the form
\eqn\expint{Ei(x)=-\int_{-x}^{\infty}{e^{-t}\over t} dt}
for $x<0$ with the general case defined by analytic 
continuation.  Asymptotically, if ${\vec p}\cdot{\vec b}\rightarrow\infty$,
the function $g\rightarrow{e^{i{\vec p}\cdot{\vec b}\over 2}
\sin{{\vec p}\cdot{\vec b}\over 2}\over 2\pi{\vec p}\cdot{\vec b}}$.
If ${\vec p}\cdot{\vec b}=0$ and ${\vec p}\times{\vec b}\rightarrow\infty$,
$g\sim({\vec p}\times{\vec b})^{-2}$.  For $p\rightarrow 0$, $g$
diverges as $\ln{pb}$, but there is always a factor of $p^2$ so
that this part of the amplitude vanishes.

For large $b$ the dominant contributions to the amplitudes are
\eqn\ampaxdom{\eqalign{{\cal A}_{\alpha\beta}^{ax}(p)
&\sim(-1+4\delta_{\alpha\beta})
\alpha'\pi 
\int dx^+ dx^- (e^{i{{\vec Y}_1(x^-)\cdot{\vec p}\over 2}
+i{{\vec Y}_2(x^+)\cdot{\vec p}\over 2}-ip^+x^- -ip^- x^+})\cdot\cr
&[ia^p(1-2{\vec y}_+\cdot{\vec y}_-){p^2\over 2}
{e^{-i{\vec p}\cdot{\vec b}\over 2}\over{\vec p}\cdot{\vec b}}]\cr}} 
\eqn\ampdildom{\eqalign{{\cal A}_{\alpha\beta}^{\phi}(p)
&\sim(-1+4\delta_{\alpha\beta})
\alpha'\pi 
\int dx^+ dx^- (e^{i{{\vec Y}_1(x^-)\cdot{\vec p}\over 2}
+i{{\vec Y}_2(x^+)\cdot{\vec p}\over 2}-ip^+x^- -ip^- x^+})\cdot\cr
&[\phi^p\{ -{p^2\over 4}
{\sin{{\vec p}\cdot{\vec b}\over 2}\over {\vec p}\cdot{\vec b}}\cr
&+({p^2\over 4}
{\sin{{\vec p}\cdot{\vec b}\over 2}\over {\vec p}\cdot{\vec b}}
-i{p^2\over 4}{e^{i{\vec p}\cdot{\vec b}\over 2}\over{\vec p}\cdot{\vec b}})
({\vec y}_+\cdot{\vec y}_-)\cr
&-({1\over 2}{\sin{{\vec p}\cdot{\vec b}\over 2}\over {\vec p}\cdot{\vec b}}
+{i\over 4}{e^{i{\vec p}\cdot{\vec b}\over 2}\over{\vec p}\cdot{\vec b}})
({\vec p}\cdot{\vec y}_+)({\vec p}\cdot{\vec y}_-)\}]\cr}}
\eqn\ampgravdom{\eqalign{{\cal A}_{\alpha\beta}^{h}(p)
&\sim(-1+4\delta_{\alpha\beta})
\alpha'\pi 
\int dx^+ dx^- (e^{i{{\vec Y}_1(x^-)\cdot{\vec p}\over 2}
+i{{\vec Y}_2(x^+)\cdot{\vec p}\over 2}-ip^+x^- -ip^- x^+})\cdot\cr
&[h_1^p\{ {ip^2\over 2}
{\sin{{\vec p}\cdot{\vec b}\over 2}\over {\vec p}\cdot{\vec b}} 
({\vec y}_+\cdot{\vec y}_-)\cr
&-({1\over 4}{\sin{{\vec p}\cdot{\vec b}\over 2}\over {\vec p}\cdot{\vec b}}
+{i\over 2}{e^{i{\vec p}\cdot{\vec b}\over 2}\over{\vec p}\cdot{\vec b}})
({\vec y}_+\times{\vec p})({\vec y}_-\cdot{\vec p})\cr
&+({3\over 4}{\sin{{\vec p}\cdot{\vec b}\over 2}\over {\vec p}\cdot{\vec b}}
+{i\over 4}{e^{i{\vec p}\cdot{\vec b}\over 2}\over{\vec p}\cdot{\vec b}})
({\vec y}_+\cdot{\vec p})({\vec y}_-\times{\vec p})\}\cr
&-h_2^p\{ -{ip^2\over 8}
{\cos{{\vec p}\cdot{\vec b}\over 2}\over {\vec p}\cdot{\vec b}}  
({\vec y}_+\cdot{\vec y}_-)
-{i\over 8}{e^{i{\vec p}\cdot{\vec b}\over 2}\over{\vec p}\cdot{\vec b}}
({\vec p}\cdot{\vec y}_+)({\vec p}\cdot{\vec y}_-)\}].\cr}}
Notice that the amplitudes vanish for unidirectional travelling waves.
The dominant radiation propagates parallel or antiparallel to ${\vec b}$.
The amplitudes for radiation transverse to ${\vec b}$ (including the
string direction) are suppressed by ${1\over b}$.
The axion amplitude is proportional to the electric field on
the D-strings.  This effect of causing axion radiation by turning
on an electric field along the string can be compared to the effect
of causing a bulk
anomaly inflow current for a chiral string by an electric field on the string.
If the fluctuations of the axion, dilaton, and graviton do not depend
on both $x^+$ and $x^-$, their Fourier transforms will
contain delta functions of $p^+$ or $p^-$, and the radiation will
be confined to the string.

Assume that the strings are far
enough apart that there is an approximate solution in supergravity
that is the sum of the right and left moving fluctuations or that there
is a linear approximation even though supersymmetry is broken.  
We assume that the above result is not greatly affected by a small
resolution of the orbifold.  The metric for the resolved orbifold
is more complicated than for the conifold but for large distances
from the D-strings, the warping factor should depend on this 
distance in the same way.  The
radiation is observed at $r_{con}=0$ (see \metric ) and 
$R>>b>>\sqrt{\alpha'}$ so we will ignore the angular dependence due
to nonzero separation.  The fluctuation of the graviton and dilaton can be 
written in position space as
\eqn\polar{e_{\mu\nu}\sim T'_{\mu\nu}(1-e^{-2\phi})\sim{T'_{\mu\nu}\over R^6}}
where the prime denotes that we drop the delta function of the 
transverse space for the energy-momentum tensor from the D-string source.
The energy-momentum tensor for the travelling wave on the D-string
was calculated from the action for the D-string in \one\ with the 
result to lowest order in $\alpha'$
\eqn\stressenergy{\eqalign{T_{+-}&=T_D(\delta^{(8)}({\vec x}-{\vec Y}_1)
+\delta^{(8)}({\vec x}-{\vec Y}_2))\cr
T_{++}&=T_D\|\partial_+{\vec Y}_2\|^2 
\delta^{(8)}({\vec x}-{\vec Y}_2)\cr
T_{--}&=T_D\|\partial_-{\vec Y}_1\|^2\delta^{(8)}({\vec x}-{\vec Y}_1)\cr
T_{+i}&=-T_D \partial_+ Y^i_2\delta^{(8)}({\vec x}-{\vec Y}_2)\cr
T_{-i}&=-T_D \partial_- Y^i_1\delta^{(8)}({\vec x}-{\vec Y}_1)\cr}}
where $T_D={1\over 2\pi\alpha' g_s}$ is the D-string tension.
In this linear approximation there is only radiation along 
the strings
because there are no terms in the energy-momentum tensor containing
an interaction of the two pulses, and the Fourier transform of the
fields will contain delta functions of $p^+$ or $p^-$.
There will be higher order terms with 
these interactions leading to gravitational radiation transverse to
the strings,
but the calculation will break down for $b\rightarrow b_{crit}$.

To obtain the second order of the graviton fluctuation is difficult.
There are $\alpha'$ corrections to the D-string action and to the 
supergravity fields which couple to this action.  In section three
we have shown that there is a potential
for the nonsupersymmetric interaction proportional to $
{\vec y}_+\cdot{\vec y}_-$.  For $b>>b_{crit}$,
this interaction is of order $1\over b^2$.  To second order the
energy-momentum tensor calculated from the D-string action by itself
will not
be covariantly conserved because one has to include the energy-momentum
of the radiation in the bulk.
Rather than trying to determine
precisely the second order fluctuation we will assume that a conserved
energy-momentum tensor of the full theory incorporating the nonlocal 
interaction can be defined where we still take $b>>b_{crit}$ so that we are in
a weak gravity limit and can use a linear approximation.  
This energy-momentum tensor takes the form
\eqn\stress{T_{\mu\nu}=T^{(4)}_{\mu\nu}\delta^{(6)}({\vec x})}
where $\mu\in\{ 0,1,2,3\}$, and we are ignoring in this formula a
small amount of smearing on the resolved orbifold.  
We choose a gauge such that the
three degrees of freedom of the graviton and dilaton satisfy
$(\bigtriangleup^{(9)}-\partial_t^2)e_{\mu\nu}=
(2\pi{\alpha'})^{5\over 2}g_s T_{\mu\nu}$ 
where $\bigtriangleup^{(9)}$
is the Laplace operator in nine dimensions.  The factor of 
$(2\pi{\alpha'})^{5\over 2}$ is introduced so that the vertex operator
is dimensionless in target space units.
This calculation should be taken 
with a grain of salt.  We only need to know $e_{\mu\nu}$ far from
the source in the Minkowski dimensions.  Assuming as we are that
there is no angular dependence in the extra dimensions, we
should be able to approximate the result by using the flat space
Laplacian.  The retarded Green function
for the ten-dimensional wave equation takes the form
\eqn\green{G(x, x')={-1\over 7\Omega_8 r^7}
(\delta(\bigtriangleup_{x,x'})-r\delta'(\bigtriangleup_{x,x'})
+{2\over 5}r^2\delta''(\bigtriangleup_{x,x'})-
{1\over 15}r^3\delta'''(\bigtriangleup_{x,x'}))}
where $\bigtriangleup_{x,x'}=t'-t+|{\vec x}_{9}-{\vec x'}_{9}|$,
$r=|{\vec x}_{9}-{\vec x'}_{9}|$, and $\Omega_8={32\pi^4\over 105}$.

We have seen that if the source contains a noninteracting piece for
large separations, the radiation is confined to the string.
In what follows we assume that the source is interacting.
If we consider finite energy pulses as
physically sensible, then the source for the radiation is not the
entire worldvolume of the D-strings but only the localized regions where both
pulses are nonvanishing.  We assume that the spatial regions can be
enclosed in a compact region of radius $r_s<<r$ where $r$ is the
observer's radius and $r_s\approx b$.  
Note that the observer can be along the string 
far from the source.
The dilaton and graviton  
satisfy 
\eqn\dilgrav{\eqalign{e_{\mu\nu}({\vec x},t)&\sim 
{(2\pi{\alpha'})^{5\over 2}g_s\over 7\Omega_8}\int d^3 x' 
{1\over |{\vec x}-{\vec x'}|^7}
(T^{(4)}_{\mu\nu}({\vec x'},t-|{\vec x}-{\vec x'}|)
-|{\vec x}-{\vec x'}|\partial_t 
T^{(4)}_{\mu\nu}({\vec x'},t-|{\vec x}-{\vec x'}|)
\cr &+{2\over 5}|{\vec x}-{\vec x'}|^2
\partial^2_t T^{(4)}_{\mu\nu}({\vec x'},t-|{\vec x}-{\vec x'}|)
-{1\over 15}|{\vec x}-{\vec x'}|^3 \partial^3_t 
T^{(4)}_{\mu\nu}({\vec x'},t-|{\vec x}-{\vec x'}|))\cr}}
Taking the Fourier transform of $T_{\mu\nu}$ in the time direction
yields
\eqn\stressfourier{T^{(4)}_{\mu\nu}({\vec x},t)={1\over \sqrt{2\pi}}
\int_{-\infty}^{\infty}d\omega 
T^{(4)}_{\mu\nu}({\vec x},\omega)
e^{-i\omega t}}
where the origin of the coordinate ${\vec x}$ lies at the center of
the compact region and at the (resolved) orbifold point,
$\omega={p^+ +p^-\over\sqrt{2}}$,
and 
$T_{\mu\nu}^{*(4)}({\vec x},-\omega)=
T_{\mu\nu}^{(4)}({\vec x},\omega)$.  Note that the coordinate ${\vec x}$
includes the extra dimensions (these are zero for $T_{\mu\nu}^{(4)}$), 
but our observations will always be
at the (resolved) orbifold point.
Then
\eqn\gravdiltwo{\eqalign{e_{\mu\nu}({\vec x},t)&\sim 
{4\pi^2{\alpha'}^{5\over 2}g_s\over 7\Omega_8}
\int_{-\infty}^{\infty}d\omega e^{-i\omega t}
\int {d{\vec x}' \over |{\vec x}-{\vec x'}|^7}
e^{i(\omega |{\vec x}-{\vec x'}|)}
T^{(4)}_{\mu\nu}({\vec x'},\omega)\cdot\cr
&(1+i\omega |{\vec x}-{\vec x'}|
-{2\over 5}\omega^2 |{\vec x}-{\vec x'}|^2 
+{1\over 15}i\omega^3 |{\vec x}-{\vec x'}|^3).\cr}}
In the limit
that the observation distance $r$ is much larger than the extent
of the source, we can approximate $|{\vec x}-{\vec x'}|$
by $r-{{\vec x}\cdot{\vec x'}\over r}$.  Thus, we obtain
\eqn\farfieldapprox{\eqalign{e_{\mu\nu}({\vec x},t)&\sim 
{i{\alpha'}^{5\over 2}g_s
\over 8\pi^2 r^4}\int_{-\infty}^{\infty}
d\omega  \omega^3 e^{-i\omega t+i{\vec k}\cdot
{\vec x}}
\int d{\vec x'}
e^{-i{\vec k}\cdot{\vec x'}}T^{(4)}_{\mu\nu}({\vec x'},\omega)\cr}}
where ${\vec k}={\omega{\vec x}\over r}$.
Obviously, there is a divergence for travelling waves that are
highly localized in the lightcone directions so for this model, we must
consider waves with compact support in $\omega$ or, alternatively,
temporally periodic waves with discrete frequencies.  For periodic waves we
would calculate the power rather than the energy.

The polarization tensor of the graviton and dilaton are obtained
from the above equation as
\eqn\fourier{e_{\mu\nu}({\vec x},\omega)=
{i{\alpha'}^{5\over 2}\omega^3 g_s
\over 2 r^4}
T^{(4)}_{\mu\nu}({\vec k},\omega).}
From this formula we can extract the forms of $\phi^p$, $h_1^p$,
and $h_2^p$ at large distances $r$.
\eqn\farforms{\eqalign{\phi^p
&\sim{i{\alpha'}^{5\over 2}\omega^3 g_s\over 2 r^4}
\phi({\vec k},\omega )\cr
h_1^p&\sim{i{\alpha'}^{5\over 2}\omega^3 g_s\over 2 r^4}
h_1({\vec k},\omega )\cr 
h_2^p&\sim{i{\alpha'}^{5\over 2}\omega^3 g_s\over 2 r^4}
h_2({\vec k},\omega )\cr}}
where 
\eqn\phihonehtwo{\eqalign{\phi({\vec k},\omega)&=
{1\over 2}\eta^{\mu\nu}T^{(4)}_{\mu\nu}( {\vec k},\omega)\cr
h_1({\vec k},\omega )&={p^+{\vec T}_+^{(4)}\times{\vec p}-
p^-{\vec T}_-^{(4)}\times{\vec p}\over p^2}\cr
h_2({\vec k},\omega )&=2
{p^+{\vec T}_+^{(4)}\cdot{\vec p}-
p^-{\vec T}_-^{(4)}\cdot{\vec p}\over p^2}\cr}}
with ${\vec k}=({p^+ -p^-\over\sqrt{2}},p^2,p^3)$ and 
${\vec T}_+^{(4)}\times{\vec p}=T^{(4)}_{+i}p_j\epsilon^{ij}$. The
condition $2p^+ p^-=p^2$ is imposed by hand.  Note again
that $h_2$ and $\phi$ are not gauge invariant, but we replace them
in the gauge we have chosen by the right hand side of the above formulas 
which is gauge invariant.  The boundary states break Lorentz invariance.

We now calculate the energy radiated through a large sphere of radius
$r$ surrounding the interaction region of the D-strings.  
Not all of the radiation is in
four dimensions--there is also radiation in the extra dimensions.
To not violate energy conservation we need to consider an eight-sphere
of radius $r$ surrounding the source.  Then we set all but two of the
angles on the eight-sphere to ${\pi\over 2}$ so that we observe the 
radiation through a two-sphere at the (resolved) orbifold point.  Again,
we are assuming no angular dependence in the extra dimensions.  
The energy flux per unit solid angle observed at the
(resolved) orbifold point is given by
\eqn\energy{{dE\over d{\Omega_8}}=r^8\int_0^{\infty}d\omega
\omega^8|{\cal A}|^2}
where $\cal A$ is
the amplitude from the string calculation.  
At large distances energy is conserved because there
is no $r$ dependence, and we obtain
\eqn\endil{{dE_{\phi}\over  d{\Omega_8}}
\sim {g_s^2{\alpha'}^{ 7}\over 4} \int_0^{\infty}d\omega
\omega^{14}|f^{\phi}\phi|^2 }
\eqn\engrav{{dE_{h}\over   d{\Omega_8}}
\sim {g_s^2{\alpha'}^{ 7}\over 4} \int_0^{\infty}d\omega
\omega^{14}(|f^{h_1}h_1|^2+|f^{h_2}h_2|^2)}
where $f^{\phi}={{\cal A}^{\phi}(p)\over\phi^p\alpha'}$,
$f^{h_1}={{\cal A}^{h_1}(p)\over h_1^p \alpha'}$,
and $f^{h_2}={{\cal A}^{h_2}(p)\over h_2^p\alpha'}$.

The factors $f^{\phi}$, $f^{h_1}$, and $f^{h_2}$  greatly
restrict the range of $\omega$ for slowly varying pulses so that these 
integrals should converge despite the enormous powers of frequency.
In the field theory limit we might expect a suppression of short
wavelengths relative to the separation of the strings.
We find a suppression of order $|g|^2|\phi|^2\sim{\sin^2{(bk)}\over (bk)^6}$
(since the amplitude in section 3 is $\sim{1\over b^2}$ and $\phi$ is
a dimensionless function of ${\vec k}$, $\phi\sim{1\over b^2 k^2}$)
so the suppression is a power rather than exponential.
Radiation with the wave vector transverse to the separation
direction ${\vec b}$ is suppressed by an additional 
factor of ${1\over b^2 k^2}$
relative to radiation parallel to this direction.
For the pulses to have finite energy $y_+$ and $y_-$ must
decrease to zero for large $|x^+|$ and $|x^-|$, and the higher derivatives
are responsible for the amplitudes ${\cal A}(p)$ not vanishing.
For more intense pulses our approximations break down, and these
integrals will diverge.  Reducing the number of dimensions reduces
the possible frequency divergence.  For
instance, for six noncompact dimensions, ignoring Kaluza-Klein
effects, one would obtain
\eqn\energysix{{dE^{six}_{\phi}\over  d{\Omega_4}}\sim 
{g_s^2{\alpha'}^{3}\over 4}
\int_0^{\infty}d\omega
\omega^{6}f^{\phi}_6|\phi|^2 }
Note that had we taken spatially periodic waves with finite energy 
per unit length,
the source of radiation would be the entire string not just localized pieces
and the calculation would be modified as in Reference \twelve .

\newsec{Conclusions}

Let us summarize our results.  We have constructed supersymmetric
boundary states
for fractional D-strings with travelling waves on a ${\bf C}^3/
{{\bf Z}_2\times {\bf Z}_2}$ orbifold.  The underlying 
orbifold with D-strings is supersymmetric.  The interaction between
two of these D-strings with waves travelling in opposite directions 
breaks supersymmetry  and was calculated.
Whether the interaction was attractive or repulsive depended
on the sign of ${\vec y}_+\cdot{\vec y}_-$ as well as whether or
not the fractional branes were of the same type.  This interaction
was localized on the D-strings to the regions where 
${\vec y}_+\cdot{\vec y}_-$ was nonvanishing and could change sign
along the D-string worldvolume.  In all cases a tachyon developed
at small separation of the two strings as an open string mode
became massless in the region where the
interaction was nonvanishing.  One might expect that there 
would be a tachyonic 
condensation with antiparallel momentum modes annihilating.
The end result would be a BPS state with momentum in one
direction.  The tachyon is related to 
massless modes in the RR sector.  The
relevant term in the supergravity would be of the form
$\int H^{RR}_+\cdot H^{RR}_-$ where $H^{RR}_{\pm}$ is the 
three-form RR field strength due to the D-string with a right
or left-moving pulse.  We found that for small separations of
the D-strings such that the tachyon was present, supersymmetry
could not be restored continuously.  This result was physically
reasonable since the target space supersymmetry was chiral.

Under the assumption that for large separations of the D-strings
compared to the critical separation, perturbative string theory
was not entirely invalid, we calculated the axion, dilaton, and graviton 
radiation emitted from the interaction of antiparallel
pulses.  By a gauge transformation one can consider the axion and
dilaton as part of the gravitational field.
At large enough separations the effective energy-momentum tensor
does not incorporate the nonlocal interaction between the two pulses,
and the radiation is restricted to the D-strings.   
The gravitational radiation
is the result of left and right moving momenta annihilating
into gravitons.  This interaction deforms the waves over time
so that one must somehow obtain an energy-momentum tensor that
incorporates the D-strings as well as the bulk.  We were not
able at this time to obtain the appropriate interacting energy-momentum
tensor but assuming its existence showed that there was gravitational
radiation.  There are two variables in the model governing the
intensity of the interaction.  Increasing the distance between the D-strings
decreases the interaction.  The amount of radiation can be
tuned to an arbitrarily small value by studying boundary states 
in which the frequency distribution has $|p^+|$ or $|p^-|$ very small 
so that, for example, the
pulse on the first D-string is a very slowly varying function of $x^-$,
while the pulse on the second D-string is an arbitrary function of $x^+$.
One could perhaps have enough control over the time dependent interaction
to calculate the variation of the waves leading 
to a supersymmetric configuration.  This interaction is significantly less
intense than brane-antibrane annihilation so it might be more easily
understood.  At the next order of $\alpha'$, this calculation
probably requires a fast computer.  
The discontinuity between the supersymmetric and
nonsupersymmetric configurations necessitates a nonperturbative
formulation.

There are differences between this string theory calculation and the
classical gravity calculation of cosmic string radiation as in \twelve .
We are required by supersymmetry to separate the right and left
moving pulses so that the interaction is nonlocal.  The situation
of having both type of pulses on the same string would not give a
controllable interaction in string theory.  How to obtain the
energy-momentum tensor for this kind of nonlocal interaction is
challenging.  We are also required to include noncompact extra
dimensions so that the metric is asymptotically flat at large
distances, and the string theory calculation makes sense.  As a result
the energy spectrum is unrealistic.  Perhaps an analogous calculation
could be done in four dimensions in an orientifolded theory.  The
time dependence of these backgrounds is somewhat trivial because
the wave only travels in one spatial dimension.  Having constructed
supersymmetric boundary states, studying waves in perpendicular 
directions could be a means for obtaining backgrounds with nontrivial
three-dimensional dependence.  The calculations in this paper were
done in the closed string boundary state framework.  One could
possibly study these questions from the dual open string viewpoint
via the S-matrix formalism developed in \one .  One still needs to
understand the nonperturbative dynamics of supersymmetry breaking.

\bigskip\centerline{\bf Acknowledgments}\nobreak
This paper and my life are dedicated to the many courageous and
talented women who have been very effectively murdered by sexual
harassment.  May we rest in peace.  May those who remain learn a
lesson from our lives.
I wish to thank J. Distler, W. Fischler, and A. Loewy for 
helpful discussions.  This work was supported in part
by the National Science Foundation under Grant No. 0071512.

\listrefs

\end